\documentclass[fleqn,usenatbib,useAMS]{mnras}

\usepackage{graphicx}	
\usepackage{amsmath}	
\usepackage{amssymb}	
\usepackage{multicol}        
\usepackage{bm}		
\usepackage{pdflscape}	
\usepackage{lipsum}
\usepackage{subcaption}
\usepackage{float}
\usepackage{longtable}
\usepackage[utf8]{inputenc}
\usepackage{comment}
\usepackage{lineno}

\usepackage{eso-pic}

\AddToShipoutPictureBG*{%
  \AtPageUpperLeft{%
    \hspace{0.75\paperwidth}%
    \raisebox{-3.5\baselineskip}{%
      \makebox[0pt][l]{\textnormal{DES-2020-0582}}
}}}%

\AddToShipoutPictureBG*{%
  \AtPageUpperLeft{%
    \hspace{0.75\paperwidth}%
    \raisebox{-4.5\baselineskip}{%
      \makebox[0pt][l]{\textnormal{FERMILAB-PUB-21-288-AE}}
}}}%

\newcommand{\kms}{\,km\,s$^{-1}$} 

\usepackage[T1]{fontenc}
\usepackage{ae,aecompl}

\usepackage{newtxtext,newtxmath}

\title{Velocity Dispersions of Clusters in the Dark Energy Survey Y3 redMaPPer Catalog}

\author[DES Collaboration]{
\parbox{\textwidth}{
  \Large
V. Wetzell,$^{1,2,3}$\thanks{Contact e-mail: \href{mailto:vwetzell@sas.upenn.edu}{vwetzell@sas.upenn.edu}}
T.E. Jeltema,$^{1,2}$\thanks{Contact e-mail: \href{mailto:tesla@ucsc.edu}{tesla@ucsc.edu}}
B. Hegland,$^{1,2}$
S. Everett,$^{1,2}$
P.A. Giles,$^{4}$
R. Wilkinson$^{4}$,
A. Farahi,$^{5,6}$
M. Costanzi$^{7,8,9}$, 
D.L. Hollowood$^{1,2}$, E. Upsdell$^{4}$,  A. Saro$^{7,8,9,10}$, 
J. Myles$^{11,12,13}$,
A. Bermeo$^{4}$,
S. Bhargava$^{4}$,
C.A. Collins$^{14}$, 
D. Cross$^{1,2}$, 
O. Eiger$^{1,2}$, 
G. Gardner$^{1,2}$,  M. Hilton$^{15}$,
J. Jobel$^{1,2}$,
P. Kelly$^{1,2}$,
D. Laubner$^{1,2}$,
A.R. Liddle$^{16}$, 
R.G. Mann$^{17}$,
V. Martinez$^{1,2}$, J. Mayers$^{4}$,
A. McDaniel$^{1,2, 18}$,
A.K. Romer$^{4}$,
P. Rooney$^{4}$, 
M. Sahlen$^{19}$, 
J. Stott$^{20}$,
A. Swart$^{1,2}$, D.J.~Turner$^{4}$, P.T.P.~Viana$^{21,66}$,
T.~M.~C.~Abbott,$^{23}$
M.~Aguena,$^{24}$
S.~Allam,$^{25}$
F.~Andrade-Oliveira,$^{26,24}$
J.~Annis,$^{25}$
J.~Asorey,$^{27}$
E.~Bertin,$^{28,29}$
D.~L.~Burke,$^{12,13}$
J.~Calcino,$^{30}$
A.~Carnero~Rosell,$^{31,24,32}$
D.~Carollo,$^{33}$
M.~Carrasco~Kind,$^{34,35}$
J.~Carretero,$^{36}$
A.~Choi,$^{37}$
M.~Crocce,$^{38,39}$
L.~N.~da Costa,$^{24,40}$
M.~E.~S.~Pereira,$^{41}$
T.~M.~Davis,$^{30}$
J.~De~Vicente,$^{27}$
S.~Desai,$^{42}$
H.~T.~Diehl,$^{25}$
J.~P.~Dietrich,$^{43}$
P.~Doel,$^{44}$
A.~E.~Evrard,$^{45,41}$
I.~Ferrero,$^{46}$
P.~Fosalba,$^{38,39}$
J.~Frieman,$^{25,47}$
J.~Garc\'ia-Bellido,$^{48}$
E.~Gaztanaga,$^{38,39}$
K.~Glazebrook,$^{49}$
D.~Gruen,$^{43}$
R.~A.~Gruendl,$^{34,35}$
J.~Gschwend,$^{24,40}$
G.~Gutierrez,$^{25}$
S.~R.~Hinton,$^{30}$
K.~Honscheid,$^{37,50}$
D.~J.~James,$^{51}$
K.~Kuehn,$^{52,53}$
N.~Kuropatkin,$^{25}$
O.~Lahav,$^{44}$
G.~F.~Lewis,$^{54}$
C.~Lidman,$^{55,56}$
M.~Lima,$^{57,24}$
M.~A.~G.~Maia,$^{24,40}$
J.~L.~Marshall,$^{17}$
P.~Melchior,$^{61}$
F.~Menanteau,$^{34,35}$
R.~Miquel,$^{58,36}$
R.~Morgan,$^{59}$
A.~Palmese,$^{25,47}$
F.~Paz-Chinch\'{o}n,$^{34,60}$
A.~A.~Plazas~Malag\'on,$^{61}$
E.~Sanchez,$^{27}$
V.~Scarpine,$^{25}$
S.~Serrano,$^{38,39}$
I.~Sevilla-Noarbe,$^{27}$
M.~Smith,$^{62}$
M.~Soares-Santos,$^{41}$
E.~Suchyta,$^{20}$
G.~Tarle,$^{41}$
D.~Thomas,$^{63}$
B.~E.~Tucker,$^{56}$
D.~L.~Tucker,$^{25}$
T.~N.~Varga,$^{64,65}$
J.~Weller,$^{64,65}$
\begin{center} (DES Collaboration) \end{center}
}}

\date{Last updated \today}

\pubyear{\today}

\begin{document}
\label{firstpage}
\pagerange{\pageref{firstpage}--\pageref{lastpage}}
\maketitle

\begin{abstract}
We measure the velocity dispersions of clusters of galaxies selected by the redMaPPer algorithm in the first three years of data from the Dark Energy Survey (DES), allowing us to probe cluster selection and richness estimation, $\lambda$, in light of cluster dynamics.  Our sample consists of 126 clusters with sufficient spectroscopy for individual velocity dispersion estimates.  We examine the correlations between cluster velocity dispersion, richness, X-ray temperature and luminosity as well as central galaxy velocity offsets.  The velocity dispersion-richness relation exhibits a bimodal distribution.  The majority of clusters follow scaling relations between velocity dispersion, richness, and X-ray properties similar to those found for previous samples; however, there is a significant population of clusters with velocity dispersions which are high for their richness.  These clusters account for roughly 22\% of the $\lambda < 70$ systems in our sample, but more than half (55\%) of $\lambda < 70$ clusters at $z>0.5$.  A couple of these systems are hot and X-ray bright as expected for massive clusters with richnesses that appear to have been underestimated, but most appear to have high velocity dispersions for their X-ray properties likely due to line-of-sight structure.  These results suggest that projection effects contribute significantly to redMaPPer selection, particularly at higher redshifts and lower richnesses.  The redMaPPer determined richnesses for the velocity dispersion outliers are consistent with their X-ray properties, but several are X-ray undetected and deeper data is needed to understand their nature. 
\end{abstract}

\begin{keywords}
galaxies: clusters: general -- X-rays: galaxies: clusters
\end{keywords}



\begingroup
\let\clearpage\relax
\endgroup
\newpage

\section{Introduction}
\label{sec:1}

The growth rate of clusters of galaxies is in principle a highly sensitive probe of dark energy given that the cluster mass function is exponentially sensitive to the underlying cosmology.  In fact, cluster studies have resulted in stringent constraints on the matter density, amplitude of perturbations ($\sigma_8$), and competitive constraints on the present day dark energy density \citep[e.g.][]{Vikhlinin09, Mantz10, Mantz15, Rozo10, spt1, planck1, spt2, DESKP}.

Currently, the largest cluster samples are drawn from wide area, optical imaging surveys using color-based \citep[e.g. red-sequence;][]{Gladders, maxBCG, Murphy, redmapperI, Oguri, Licitra} or photometric redshift-based selection \citep{Dong, Milkeraitis, Durret, Soares-Santos, amico, Aguena}.  The statistical power of these cluster samples gives them the potential to be the single most constraining probe of dark energy in large-area surveys like the Dark Energy Survey \citep[DES;][]{Weinberg13,DESKP}; however, the constraining power is currently limited by systematics in cluster selection and mass calibration \citep{DESKP}.  In particular, photometric cluster selection inevitably suffers from the projection of structure along the line of sight with galaxies over a large range of distances potentially being counted as cluster members \citep[e.g.][]{Lucey83, Costanzi19a}. Spectroscopy, where available, allows for a more robust determination of cluster membership, and the velocity dispersion of member galaxies correlates with cluster mass, allowing for the calibration of some of the systematics affecting optical cluster selection \citep{redmapperIV, Farahi16, Myles}.

In this paper, we study the kinematics of redMaPPer \citep[red-sequence Matched-filter Probabilistic Percolation,][]{redmapperI,redmapperSV} selected clusters from the first three years of Dark Energy Survey data using archival spectroscopy.  Specifically, we determine the velocity dispersions of 126 clusters with at least 15 spectroscopic member galaxies and investigate the velocity dispersion-richness relation; the scatter and redshift dependence of this relation give us an indication of the types of systems selected by the cluster finding algorithm. We also look at the correlation of velocity dispersion with X-ray cluster properties where available. This study extends the examination of redMaPPer cluster selection and dynamics to higher redshifts than previous spectroscopic studies of SDSS clusters \citep{redmapperIV, Farahi16, Myles}.

The structure of this paper is as follows. In Section \ref{sec:2}, we present our cluster selection and available spectroscopy and X-ray data. In Section \ref{sec:3}, we outline the statistical methodology used to obtain velocity dispersion estimates. In Section \ref{sec:4}, we examine the velocity dispersions, the velocity disperion-richness relation, and the distribution of redMAPPer determined central galaxy velocities. In Section \ref{sec:5}, we investigate the bi-modal velocity dispersion-richness distribution in relation to other cluster properties like redshift and X-ray emission. In Section \ref{sec:6}, we summarize our findings and discuss future work.

\section{Data} 
\label{sec:2}

\subsection{Cluster Catalog}

We study the properties of clusters selected from the wide-area, optical imaging data of the Dark Energy Survey (DES) \citep{DES2005}.  Specifically, clusters are selected from the DES Y3 Gold catalog \citep{Gold} which includes data taken from the first three years of the survey covering 4946 deg$^2$ in $griz$.  These data represent a large increase in area by a factor of $\sim 2.7$ with only a modest increase in depth compared to DES Y1.

Clusters are identified in DES data using the redMaPPer algorithm, a photometric red-sequence cluster finder \citep{redmapperI,redmapperSV}.  RedMaPPer iteratively selects red-sequence galaxies and assigns them a probability of membership to clusters based on a matched filter on color, magnitude, and spatial separation from the most likely identified central cluster galaxy. An observable proxy for cluster mass is the redMaPPer determined richness, $\lambda$, which is the sum of the galaxy membership probabilities in a given cluster within a given radius \citep{redmapperI,mcclintock19}.

The data set used in this study is comprised of galaxy clusters and their respective member galaxies selected using redMaPPer version 6.4.22+2 from the DES Y3 Gold catalog.  Specifically, we consider the richness greater than 20, full cluster catalog and the associated member catalog. We will also examine results for the volume-limited, $\lambda>20$ catalog, which only includes clusters that have been observed with sufficient depth to detect the faintest galaxies used in the richness calculation, $0.2L_*$ galaxies. 

In this work, we focus on the subset of redMaPPer clusters with sufficient spectroscopy of cluster member galaxies for statistical analysis, as described below.

\subsection{Spectroscopic Catalog}

The redMaPPer member catalog includes spectroscopic redshift measurements of cluster member galaxies from archival surveys including SDSS DR14 \citep{SDSS} and the OzDES Global Redshift Catalog, which collates spectroscopy taken by the OzDES survey \citep{ozdes, Lidman} as well as data from other published spectroscopic surveys in the DES supernova fields. In addition to redshifts in the redMaPPer catalog, we included spectra from additional archival surveys as collated for DES photometric redshift calibration \citep{portal}.

As we wish to measure the peculiar velocity distributions within our clusters and to robustly probe cluster membership, we limit our sample to clusters with spectroscopic redshifts for at least 15 galaxies identified by redMaPPer as possible cluster members.  The choice of 15 as a minimum is somewhat arbitrary. A minimum sample of 10 galaxies is typically recommended for the velocity dispersion estimators we use \citep[e.g.][]{Beers}; however, the scatter in velocity dispersion estimates decrease as the number of members increase, and few-member velocity dispersions based on primarily the brightest galaxies can be biased \citep[e.g.][]{Saro13}. The minimum of 15 is chosen to strike a balance between reducing scatter and bias while not overly restricting the sample size.

As detailed in section \ref{sec:3}, after a first pass at determining the cluster central redshift, we further cull the galaxy catalog removing galaxies whose velocity offsets indicate they are not cluster members. After this cut and again requiring spectroscopic redshifts for at least 15 member galaxies, we get a final sample of 126 clusters for our analysis; of these, 76 clusters have spectra for at least 20 members. We chose not to remove member galaxies based on their redMaPPer assigned probability of membership ($P_{\rm MEM}$), as it severely limited our sample without significantly reducing the ratio of outlier clusters, as shown in Appendix \ref{sec:appendix}. It is also important to note that we do not consider any bias due to selection effects such as targeting strategies, as these spectroscopic measurements are largely archival. Ongoing programs are collecting new spectroscopy for subsets of redMaPPer clusters, which will be the subject of future work.

\subsection{X-ray data}

A number of the clusters in our sample have existing X-ray data to which we compare the velocity dispersions in Section 5.  Out of our total sample of 126 clusters, 30 have archival Chandra observations and 43 have archival XMM-Newton observations, after removing clusters where the proximity to the detector edge or other clusters prevented accurate analysis.  Eleven clusters appeared in both the Chandra and XMM samples.  For these systems, we use the XMM measurements, because the temperatures typically had smaller uncertainties.  In total, this gives a sample of 62 unique clusters with X-ray data, roughly half of our sample.

These data were reduced and analyzed with the MATCha \citep{matcha} and XCS \citep{XCS,Giles} pipelines for Chandra and XMM data, respectively.  For clusters with sufficient data, the X-ray temperature and luminosity were determined through fits to the X-ray spectrum.  In this work, we utilize temperatures and luminosities within an $r_{2500}$ radius.  For X-ray detected clusters with insufficient statistics to fit the temperature, the luminosity was estimated starting with an assumed temperature of 3 keV and then iterating over the $L_X-T_X$ relation for redMaPPer clusters \citep{matcha}.  For undetected clusters we estimated the $3 \sigma$ upper limit on $L_X$ given the detected count rate in a 500 kpc aperture surrounding the redMaPPer position.  All X-ray to redMaPPer matches were visually examined and compared to known clusters and other nearby redMaPPer clusters. In some cases, the X-ray cluster was a known cluster at a different redshift and not the redMaPPer cluster being considered; these were removed from the sample. In general, given proximity, redshift, and richness the X-ray associations were unambiguous.  For details see \cite{matcha} and \cite{Giles}.

\section{Methodology}
\label{sec:3}

In this section, we outline the methods used to determine cluster redshifts and velocity dispersions for the spectroscopic sample.

\subsection{Member Selection}

Using the cluster redshift, determined by the biweight location estimator (Section \ref{Biweight Location Estimator}), we computed the peculiar velocities of spectroscopically measured galaxies which redMaPPer determined to be potential cluster members,
\begin{equation}
    v = c\ \frac{z_i-C_{BI}}{1+C_{BI}},
\end{equation}
where $c$ is the speed of light in \kms, $z_i$ is the galaxy spectroscopic redshift, and $C_{BI}$ is the cluster redshift estimated using the biweight location estimate (see Section \ref{Biweight Location Estimator}).

\begin{figure}
    \centering
    \includegraphics[width=1\columnwidth]{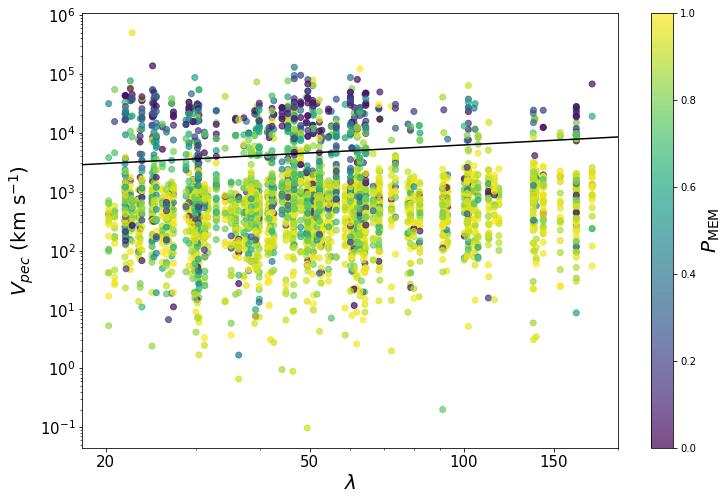}
    \caption{Member galaxy peculiar velocities shown against their cluster's richness with the galaxy probability of membership on the color axis. The black line shows the initial cut to remove interlopers from Equation \ref{eq:cut}. Color represents membership probability, $P_{\rm MEM}$ from the redMaPPer algorithm \citep{redmapperI}.}
    \label{fig:vel_offset}
\end{figure}

After determining the peculiar velocities of the potential member galaxies from redMaPPer, we make a cut on velocity offset as a first cut to remove interlopers in the foreground or background that are not cluster members. For this cut we follow the richness dependent cut presented in \cite{redmapperIV}
\begin{equation}
    |v|\leq(3000\textrm{ km s}^{-1})\left(\frac{\lambda}{20}\right)^{0.45},
    \label{eq:cut}
\end{equation}
where $\lambda$ is the richness of the cluster to which the galaxy is a member. Figure \ref{fig:vel_offset} shows the peculiar velocities versus richness for our initial sample along with a line showing the cut for non-members. 

\subsection{Statistics}

For determination of cluster central redshift and velocity dispersion, we follow the methods detailed in \cite{Beers}.
In this section we summarize the resistant and robust location and scale estimators from \cite{Beers} utilized in this work, in particular those appropriate to the few-$N_{\rm members}$ regime of $N_{\rm members}\ge 15$ to obtain velocity dispersions for galaxy clusters that do not have complete spectroscopic sampling or have not reached dynamic equilibrium. We will specifically utilize and compare the biweight scale and gapper methods of estimating the velocity dispersion.

\subsubsection{Biweight Location Estimator} \label{Biweight Location Estimator}

The biweight location estimator is used to determine the redshift of the cluster based on the redshift of the member galaxies listed in the redMaPPer member catalog. We chose this location estimate since it is robust in the presence of non-Gaussian initial populations and resistant to contaminated normal distributions.

For a set of redshift measurements $Z$, the biweight location estimator is defined as
\begin{equation}
    C_{BI}(Z) = M + \frac{\sum_{|u_i|<1}(z_i-M)(1-u_i^2)^2}{\Sigma_{|u_i|<1}(1-u_i^2)^2},
\end{equation}
where $M$ is the sample median and $u_i$ is defined as
\begin{equation}
    u_i = \frac{(z_i-M)}{C\,MAD(z_i)}.
\end{equation}
The constant $C$ is the "tuning constant" and is set to $C=6$ for the best balance of efficiency across a broad range of initial populations, and the function $MAD(z_i)$ is the median absolute deviation of the redshifts given by
\begin{equation}
    MAD(z_i) = \textrm{median} (|z_i-M|).
    \label{eq:MAD}
\end{equation}
We iterated this process 10 times to obtain a more accurate central redshifts by setting $M$ equal to $C_{BI}$ from the previous iteration.

\subsubsection{Biweight Scale Estimator}

The biweight scale estimator is an unbiased estimator which can be used to determine velocity dispersions of galaxies within a cluster when there are few measurements. This estimator is resistant to outliers (in this case, interloping galaxies), unlike the sample mean, and is robust against variance in the assumed probabilistic model of the sample population. It is important to note that the associated variance \citep{Beers} is biased similarly to the population variance, however, the sample variance is not. Due to this we have followed the biweight scale estimate \citep{ruel_optical_2014} which is
\begin{equation}
    \sigma_{BI}^2 = N_{\rm members}\frac{\sum_{|u_i|<1}(1-u_i^2)^4(v_i-\overline{v})^2}{D(D-1)}
\end{equation}
where $v_i$ are the peculiar velocities and $\overline{v}$ is the average of the peculiar velocities. $D$ is defined as 
\begin{equation}
    D = \sum_{|u_i|<1}(1-u_i^2)(1-5u_i^2)
\end{equation}
where $u_i$ is defined as
\begin{equation}
    u_i=\frac{v_i-\overline{v}}{C\,MAD(v_i)}.
\end{equation}
The constant $C$ is once again the tuning constant which is set to $C=9$ for the scale estimator and $MAD(v_i)$ is defined similarly to Equation \ref{eq:MAD}. We iterated this process using a 3-sigma clipping to obtain a more accurate estimate by removing interlopers.

\subsubsection{The Gapper Method}

The gapper method is a scale estimator based on the gaps between ordered measurements. For the ordered measurements $v_i,v_{i+1},\ldots,v_n$, the gaps are defined as
\begin{equation}
    g_i=v_{i+1}-v_i,\ \ \ \ \ i=1,\ldots,n-1.
\end{equation}
The approximately Gaussian weights of these gaps are given by
\begin{equation}
    w_i=i(n-i).
\end{equation}
The gapper scale estimate is then defined as
\begin{equation}
    \sigma_G=\frac{\sqrt{\pi}}{n(n-1)}\sum^{n-1}_{i=1}w_ig_i.
\end{equation}
The gapper method is well adapted for our data set as it can efficiently determine accurate scale estimates for as few as $N_{\rm members}=10$ measurements without being strongly influenced by interlopers.

\subsection{Confidence Intervals}

Confidence intervals for the velocity dispersions were established using a bootstrap resampling with replacement. We created 10,000 resampled galaxy catalogs for each cluster in the study. We applied both the biweight scale estimate and gapper method to each of these resampled clusters. We chose our listed velocity dispersion to be the median measurement of the resampled clusters and set our confidence intervals to contain 68\% of the measurements around the median.

\section{Results}
\label{sec:4}

\subsection{Velocity Dispersions}

\begin{figure}
    \centering
    \includegraphics[width = 1\columnwidth]{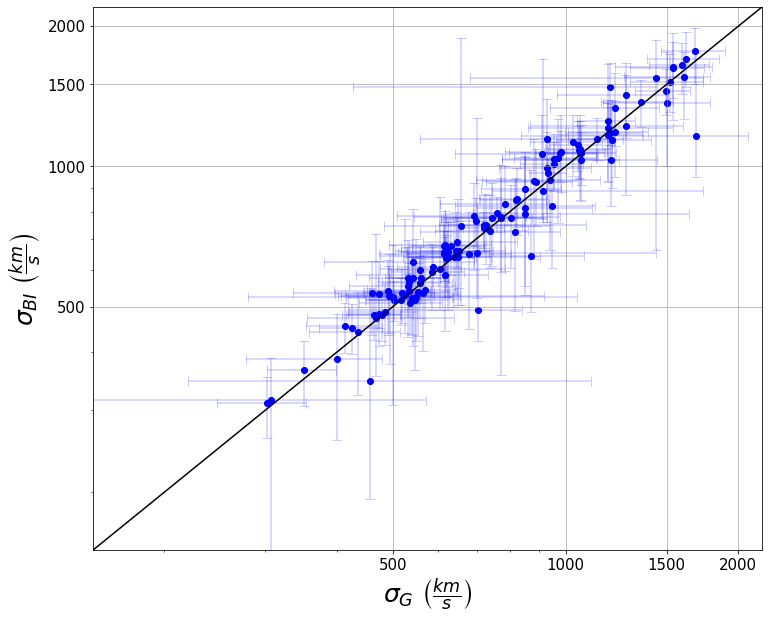}
    \caption{Comparison of the velocity dispersion estimators $\sigma_G$ and $\sigma_{BI}$. The black line shows a one to one relation between both estimators. It is apparent that these estimators agree well for the majority of the clusters in our sample. }
    \label{fig:sig_g_sig_bi}
\end{figure}

\begin{figure*}
    \centering
    \includegraphics[width = 2\columnwidth]{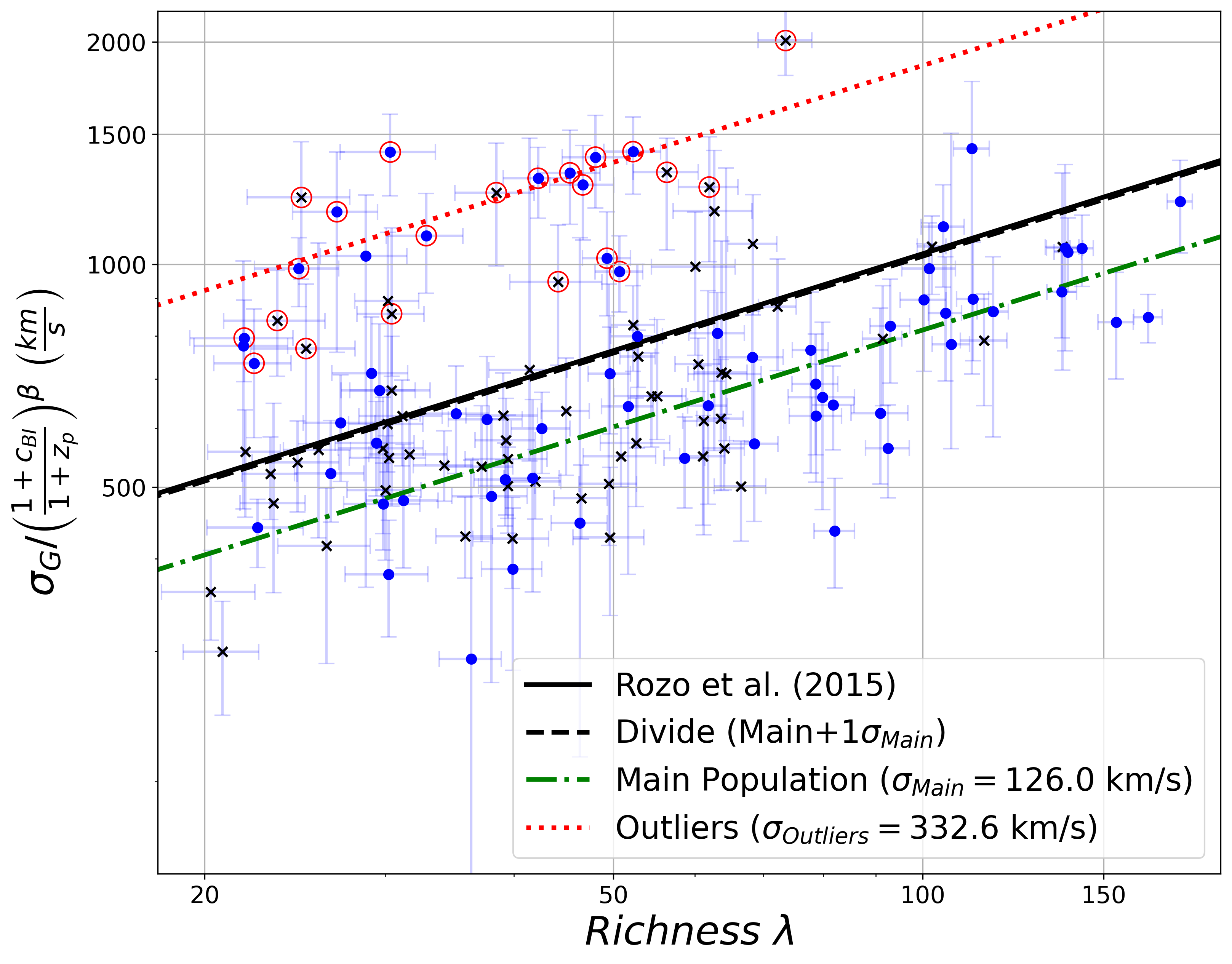}
    \caption{Velocity dispersion-richness relation for velocity dispersion estimated with the gapper method.  The black line shows the $\lambda$ adjusted $\sigma_v$--$\lambda$  trend line found by \citet{redmapperIV}, with the richness adjusted for the difference between SDSS and DES Y3 using equation \ref{eq:sdss2desy3}. The redshift dependence is accounted for by adjusting the cluster velocity dispersion based on the cluster redshift ($z_p=0.171$, $\beta=0.54$) following the redshift dependence found in \citet{redmapperIV}. The $\sigma_G$--$\lambda$ relation shows a bi-modal distribution with a small, but significant fraction of clusters having apparent velocity dispersions that are high for their richnesses. We use a double Gaussian fit to the residuals relative to the \citet{redmapperIV} line to separate the two populations giving best-fit $\sigma_G$--$\lambda$ relations for the main and outlier populations assuming the same slope as \citet{redmapperIV} but offset in normalization.  The green dash-dotted line indicates the center of the main population, while the red dotted line indicates the center of the outlier population. The black dashed line shows a 1$\sigma$ deviation from the center of the main population. We define outlier clusters to be any systems whose velocity dispersion lower limit is above the dashed black line.  Clusters that are in the volume limited catalog are marked with a black x. Clusters that are outliers are circled in red.}
    \label{fig:lamdba_sigma}
\end{figure*}

We determined the velocity dispersions of the clusters using both the biweight scale estimate and the gapper method with the results listed in Table \ref{tab:results}. We found that the biweight scale estimate agreed well with the gapper method which is apparent from both Table \ref{tab:results} and  Figure \ref{fig:sig_g_sig_bi} showing the relation between velocity dispersion estimates for the two methods. The biweight scale estimates of 125 of the 126 clusters are contained within the confidence intervals of their respective gapper scale estimates. Due to the high level of non-Gaussianity in our sample and the presence of significantly offset interlopers, we chose to focus our investigation of the velocity dispersion-richness relation on the velocity dispersions obtained using the gapper method, since it appears to be more stable than the biweight scale estimate when considering the bootstrap resampling on our data. This choice is supported by analysis of both the biweight scale estimate and gapper method with simulated clusters showing that the gapper method returns a nearly constant estimate for the cluster velocity dispersion regardless of the number of sampled galaxies \citep{Ferragamo}. The velocity dispersion-richness relation using the gapper estimates is shown in Figure \ref{fig:lamdba_sigma}.

\subsection{Velocity Dispersion - Richness Relation}

Inspection of the $\sigma_G$--$\lambda$ relation in Figure \ref{fig:lamdba_sigma} reveals a bi-modal distribution. The majority of clusters appear to follow a power law relation similar to previous determinations of the velocity dispersion-richness relation \citep[e.g.][]{redmapperIV} with a slope of $\sim 0.44$ ($\sigma$ scales as $\sim \lambda^{0.44}$, see equation 8 of \citet{redmapperIV}). A smaller, but significant, population of clusters appear to have relatively high velocity dispersions for their richnesses. 

\begin{figure}
    \centering
    \includegraphics[width = 1\columnwidth]{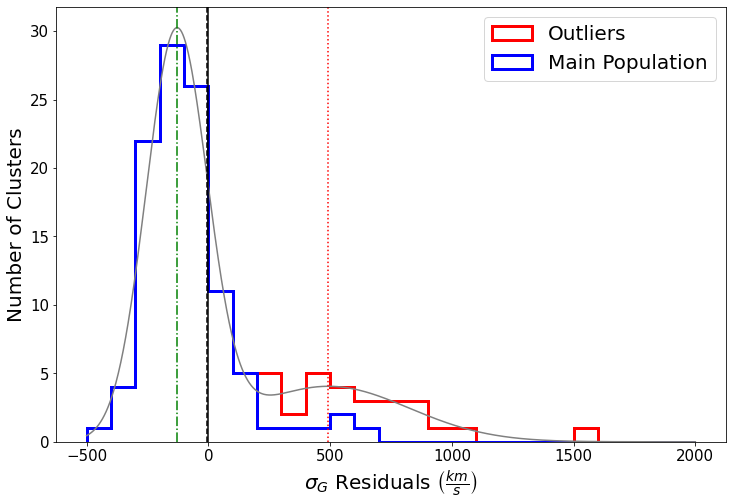}
    \caption{Stacked histogram of residuals between cluster velocity dispersions and the line found by \citet{redmapperIV}. A double Gaussian fit to this distribution is shown in grey. The center of the primary peak (representing the main cluster population) is at $-131$ \kms (green dot-dashed line) with a one-sigma width of 126 \kms. The center of the secondary peak (representing the outlier population) is at 492 \kms (red dotted line) with a width of 333 \kms. Using this information we chose to identify outliers as clusters whose velocity dispersion lower limit (68\% confidence interval) is more than one standard deviation away from the main population (black dashed line). The histogram of outlier clusters selected in this way is shown in red, and non-outliers clusters are shown in blue.}
    \label{fig:residuals}
\end{figure}

In order to separate the two populations, we examine the residuals of the cluster velocity dispersions when compared to the trend line found by \citet{redmapperIV}. We fit a double Gaussian to the residuals, as shown in Figure \ref{fig:residuals}; effectively we assume the same slope found by \citet{redmapperIV} but offset in normalization.  The main population of clusters is well fit by a Gaussian peaked at $-131$\kms relative to \citet{redmapperIV} (dot-dashed green line in Figure \ref{fig:lamdba_sigma}) with a width of 126 \kms, while the outlier population gives a secondary peak centered at 492 \kms (dotted red line in Figure \ref{fig:lamdba_sigma}) with a width of 333 \kms.  We define as outliers clusters whose lower limit on their velocity dispersion (68\% confidence interval) is more than one standard deviation from the $\sigma_G-\lambda$ relation of the main population (dashed black line in Figure \ref{fig:lamdba_sigma}). This population accounts for 17\% of the clusters in our sample; the selected outliers are circled in red in Figure \ref{fig:lamdba_sigma}.

While the slope of the $\sigma_G$--$\lambda$ relation is similar to that found in \cite{redmapperIV} for SDSS clusters, the normalization of the previous relation lies above that of our main population.  Here we have adjusted the \cite{redmapperIV} line for the difference in $\lambda$ between their SDSS sample and our DES Y3 sample.  Using clusters found in both samples, we fit for the relation between $\lambda$ in the two samples, finding
\begin{equation}
\lambda_{\rm SDSS} = (0.92\pm0.2)\lambda_{\rm DESY3} + 0.45\pm0.68
\label{eq:sdss2desy3}
\end{equation}
The sample for this comparison is limited by the overlap both in sky coverage and redshift range between SDSS and DES Y3, but in any case the offset in $\lambda$ is very small and does not affect our conclusions. 

The determination of the velocity dispersion-richness relation in \cite{redmapperIV} was based on fits to the stacked velocity offsets of pairs of galaxies, specifically the velocity offset of redMaPPer centrals from other redMaPPer member galaxies, rather than individual clusters as analyzed here.  The higher normalization may then stem from their sample containing galaxies in a mix of both typical clusters and the outlier population.  \citet{Rines18} studied the velocity dispersion - richness relation using Hectospec spectroscopy for 27 high richness, low redshift SDSS redMaPPer clusters.  Both the normalization and slope of their $\sigma-\lambda$ relation are in good agreement with what we find, though their expanded sample extending to lower richness clusters has a slope that is too shallow compared to our data \citep{Rines18}.
  
Figure \ref{fig:lamdba_sigma} shows the $\sigma_G-\lambda$ relation for clusters drawn from the "full" redMaPPer catalog with no limitation on redshift.  The redMaPPer performance is less robust at lower redshifts, due to the lack of u-band data, and at higher redshifts, due to incompleteness of the galaxy catalogs at Y3 depth. Furthermore, DES cluster cosmology studies have typically adopted a redshift range of $0.2 < z < 0.65$.  The "full" catalog, as compared to the volume-limited catalog, includes data where the local depth is not deep enough to reach the $0.2L_*$ limit used to calculate $\lambda$ and so includes clusters with extrapolation of their richnesses.  As a first test, we consider the  $\sigma_G-\lambda$ relation for only clusters in the volume-limited catalog with $0.20<z<0.65$, shown in Figure \ref{fig:lamdba_sigma} as black x's. The relation between velocity dispersion and richness including the appearance of a bimodal population is very similar, albeit with lower statistics, indicating that the outlier population does not simply stem from clusters with less robust selection compared to the core redMaPPer sample. 

Investigation of the velocity distributions for the outlier clusters reveals that they are truly broad and often non-Gaussian; only one or two show indications of a bimodal velocity distribution.  The individual and stacked histograms of all clusters in our sample are shown in Appendix \ref{sec:appendix}. In Appendix \ref{sec:appendix}, we also look at the effect of employing a more stringent initial selection of potential cluster members.  This has the effect of somewhat reducing the velocity dispersions of the outlier clusters but they still appear as a population with higher normalization in the $\sigma_G-\lambda$ relation.  This test again shows that the velocity distributions are broad and fairly continuous, not simply influenced by a small number of galaxies with large velocity offsets.

Several factors can act to inflate the observed velocity dispersion including projection effects of structure along the line of sight, the presence of substructure or correlated structures, and unremoved interloping galaxies in the foreground or background.  On the flip side, there are effects which can act to reduce the observed richness of redMaPPer-selected clusters, including miscentering and percolation \citep{Zhang19,Costanzi19a}.  If the origin of this population is related to cluster selection and characterization (e.g. projection effects, miscentering), it would have important implications for cosmological studies perhaps indicating significant richness scatter or impurity in the cluster catalog.  In Section \ref{sec:5}, we further investigate the origin of these clusters.

\subsection{Central Galaxy Velocity Distribution}

In addition to cluster velocity dispersions, we can also examine the redMaPPer redshift accuracy and the peculiar velocity distribution of the galaxies redMaPPer identifies as likely central galaxies. In this section, we examine central cluster redshifts, and we will return to examination of the velocity dispersion outliers in Section \ref{sec:5}.  

\begin{figure}
    \centering
    \includegraphics[width = 1\columnwidth]{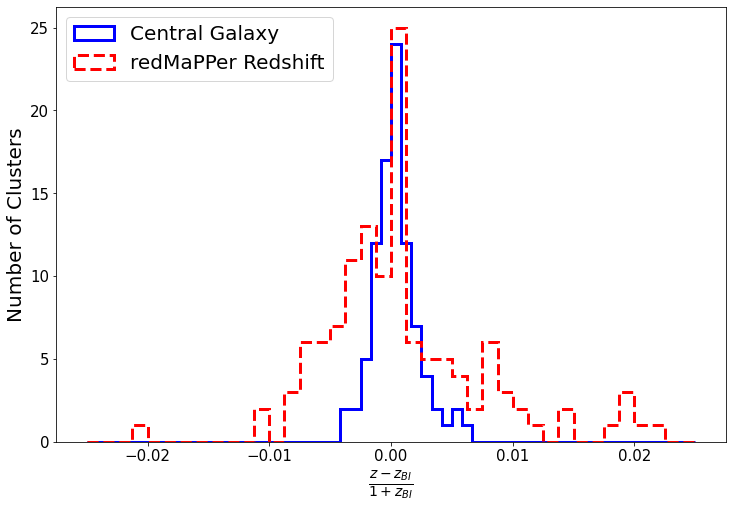}
    \caption{Difference in redshift between the biweight location estimate of the central cluster redshift and the redMaPPer estimated cluster redshift in $z=0.0013$ wide bins (dashed red). Difference in redshift between the biweight location estimate and the redMaPPer central galaxy spectroscopic redshift where available in $z=0.00083$ wide bins (solid blue).} 
    \label{fig:centers}
\end{figure}

We examined several cluster redshifts for the clusters in our study including the redMaPPer estimated redshift, the redMaPPer central galaxy redshift, and the biweight location estimate based on spectroscopic measurements.  Figure \ref{fig:centers} shows the distribution of velocity offsets between the redMaPPer estimated redshift and the biweight location for all clusters, and the distribution of velocity offsets between the central galaxy redshift and the biweight location for the 91 clusters with central galaxy spectra. The standard deviation of the redMaPPer redshifts compared to the biweight location is 0.0067. This dispersion is similar to previous determinations of the redMaPPer redshift performance; for example, \cite{mcclintock19} finds a redshift scatter, when compared to spectroscopic redshifts for the central galaxy, of $\sigma_{\rm Central\ Galaxy}/(1+z) \sim 0.006$.

Figure \ref{fig:centers} shows that the central galaxy peculiar velocities and biweight location are fairly tightly correlated with the standard deviation being 0.0018.  Nonetheless, there are putative central galaxies with velocity offsets compared to the overall cluster of up to 2000 km s$^{-1}$.  There are two likely origins of these large offsets.  The first is ongoing or recent cluster merging activity.  The second is that redMaPPer misidentified the central galaxy; miscentering by redMaPPer occurs for $\sim 20-30$\% of clusters \citep{Zhang19}.  Significant velocity offsets of centrals for clusters which are otherwise well centered and relaxed could be an indicator of self-interacting dark matter, which creates cored dark matter profiles allowing for larger oscillation of the central galaxy within the halo \citep{Kim17}.

\section{Investigation of Bi-Modal $\sigma_G$ -- $\lambda$ Populations}
\label{sec:5}

In this section, we examine the bi-modal populations of galaxy clusters in $\sigma_G$ -- $\lambda$ space. The two populations that are apparent in Figure \ref{fig:lamdba_sigma} are one, containing the majority of galaxy clusters, that has roughly the same slope as found by \cite{redmapperIV} but offset to lower velocity disperions and a smaller population with relatively low richnesses and high velocity dispersions. As previously, we define the outlier population to be clusters with confidence intervals that remain above one standard deviation from the main population.

There are a few possible origins of the outlier population.  First, they may be truly massive clusters whose richness is underestimated.  This can occur, for example, if redMaPPer significantly miscenters the cluster thus counting galaxies around the wrong location \citep{Zhang19}.  On the flip side, they may be lower mass clusters as indicated by their richness whose velocity dispersions are inflated by correlated structure (e.g. filaments, superclusters), merging activity, or unremoved interloper galaxies. In fact, using simulated clusters, \cite{Saro13} find that the interloper fraction in spectroscopic samples is expected to increase for both lower mass and higher redshift clusters as seen here, though the definition of interlopers in that work does not distinguish between contaminating galaxies in correlated structure and unrelated foreground and background galaxies.  We will use interlopers to mean a small number of unrejected background or foreground galaxies and argue that this is unlikely to be a dominant origin of the outliers, while correlated structure, galaxies in nearby superstructures or filaments, are a likely origin.

In the case of correlated structure, it is possible the observed richness is also biased high compared to the halo mass due to projection effects \citep{Costanzi19a}.  RedMaPPer down weights the membership probabilities, and therefore richness, for galaxies that are offset in color and radius from the cluster center; this weighting mitigates though does not remove the effects of projection on richness estimation \citep{Costanzi19a,Myles}. Thus, we might expect a larger bias in velocity dispersion compared to richness for crowded lines of sight. For example, a filamentary structure along the line of sight may have a very high velocity dispersion with a moderate/low richness that is nonetheless high for the true virialized mass impacting its selection.  

Understanding the nature of the velocity dispersion outliers can give us insight into the types of systems that redMaPPer selects. In the following subsections, we further examine their properties including the individual and stacked velocity distributions of these clusters (Section \ref{sec:5.1}), their spatial and redshift distributions (Section \ref{sec:5.2}), and their X-ray properties compared to the main population (Section \ref{sec:5.3}).

\subsection{Velocity Distributions and Interlopers}
\label{sec:5.1}

It is difficult from sparse spectroscopic data to entirely rule out contamination from interlopers, and these may be the cause of some of the outliers. However, a few factors argue against this being the dominant source of the outliers.  First, inspection of Figure \ref{fig:galery}, which shows the individual peculiar velocity distributions of all clusters in our sample, sheds light on the shape of the velocity distributions of the outlier population. Many of these clusters appear to have intrinsically broad distributions.  Second, cuts on galaxy membership probability (Figure \ref{fig:pmem_stacked}) or a more stringent initial cut on peculiar velocity (Figure \ref{fig:2000kms_cut}) which reduce interlopers do not significantly change the outlier population. In particular, a cut on membership probability at first appears to remove outliers (Figure \ref{fig:sig_g8_lam}); however, this was almost entirely due to individual clusters dropping below the 15 member limit for study.

\begin{figure}
    \centering
    \includegraphics[width=0.5\textwidth]{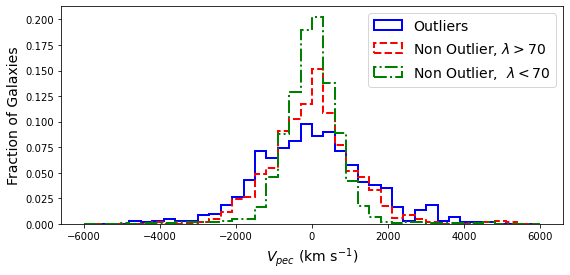}
    \caption{The fraction of galaxies in peculiar velocity bins for stacks of different cluster populations. The outlier population is shown in solid blue, the main population at high richness is shown in dashed red, and the main population at low richness is shown in dotted green.  }
    \label{fig:outlier_stacked}
\end{figure}

Figure \ref{fig:outlier_stacked} shows the stacked velocity distribution of the outlier population compared to those of rich clusters with similar velocity dispersion ($\lambda>70$ and $\sigma_G>1000$ \kms) and clusters of similar richness with low velocity dispersion ($\lambda<70$ and $\sigma_G<1000$ \kms). If the outlier clusters were simply lower mass clusters with significant contamination, we might expect to see a narrower Gaussian component, similar to other low richness clusters, plus large wings in the stacked distribution. Instead the stacked outlier population has a fat Gaussian distribution very similar to that of richer clusters. Furthermore, a Kolmogorov–Smirnov test was unable to reject the null hypothesis that the distribution of stacked member galaxies from the outlier population was drawn from the same population as the high richness, high velocity dispersion population ($\lambda>70$, $\sigma_G>1000$\kms) with a p-value of 0.45. If contamination from interlopers contributed significantly to the outlier population we would expect the stacked high richness, high velocity dispersion distribution to be markedly different than the stacked outlier distribution.

The above suggests that at least some of the outlier clusters are massive clusters which have been assigned a low richness for their mass, that these are unvirialized structures, or that they are lower mass halos living in regions with significant filamentary/correlated structure.  It is also possible that many of them are merging clusters with some line-of-sight separation which cannot be distinguished with our limited spectroscopy, but this does not appear to be the case in Figure \ref{fig:galery}.

\subsection{Projection Effects and Correlation with Redshift}
\label{sec:5.2}

An intrinsic difficulty in cluster selection from photometric data is the inability to distinguish cluster members from galaxies in projection, and galaxies $\sim100$ Mpc in front of or behind the clusters can be included by redMaPPer as potential member galaxies \citep{Costanzi19a, Sohn}.  These projection effects lead to a preferential selection of clusters with correlated structure along the line of sight \citep{DESKP, Sunayama, Wu}.  If the outlier clusters live in regions with filaments and/or supercluster environments, this could lead to the enhanced velocity dispersions, and the prevalence of these systems would tell us about the redMaPPer selection.

\begin{figure}
    \centering
    \includegraphics[width = 1.0\columnwidth]{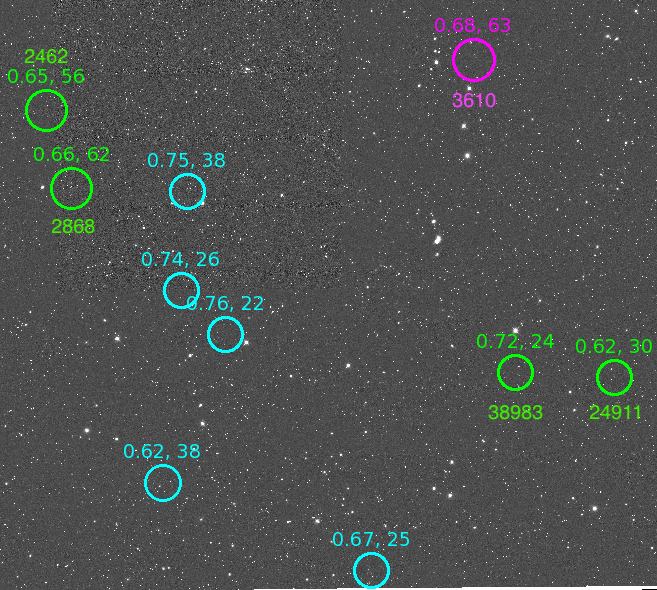}
    \includegraphics[width = 1.0\columnwidth]{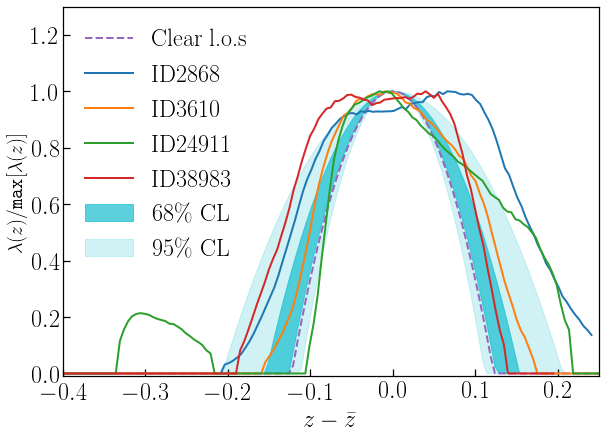}
    \caption{\textit{Top:} DES Y3A2 r-band mosaic (1.5 deg per side) of a superstructure at $z\sim0.7$.  Circles mark the positions of redMaPPer $\lambda>20$ clusters in a redshift range $0.6<z<0.8$.  Circle size indicates $R_{\lambda} = (\lambda/100)^{0.2}$ $h^{-1}$ Mpc, and region labels list $z, \lambda$; clusters in the velocity dispersion catalog are also labeled with their MEM\_MATCH\_ID. Clusters that are velocity dispersion outliers are indicated in green; MEM\_MATCH\_ID 3610 which has a high best-fit velocity dispersion but large uncertainties on $\sigma_G$ is indicated in magenta, and additional clusters at similar redshifts in cyan.  The cyan clusters do not have sufficient spectroscopy for velocity dispersion determination.  This superstructure contains at least four high velocity dispersion, low richness clusters. \textit{Bottom:} Normalized richness scanned over redshift, $\lambda(z)/$max$[\lambda(z)]$, for the outlier clusters above (green circles) and 3610 (magenta circle) which are in the volume-limited redMaPPer catalog. These are compared to the normalized $\lambda(z)$ expected for a cluster without any projection at $z=0.65$ (dashed purple line), and the 68\% and 95\% distribution of $\lambda(z)$ at the same redshift (dark and light cyan bands).}
    \label{fig:superstructure}
\end{figure}

In fact, four of the outlier clusters (MEM\_MATCH\_IDs 2462, 2868, 24911, 38983) lie within the same $\sim 2$ deg$^2$ patch of sky and within $0.1$ in redshift, as shown in Figure~\ref{fig:superstructure}.  A fifth cluster in this field and redshift range, MEM\_MATCH\_ID 3610, also has a high best-fit velocity dispersion but with large uncertainties due to a potentially bimodal velocity distribution (see Figure \ref{fig:galery}).  There are also several additional redMaPPer clusters in the same field with similar redshifts but lacking sufficient spectroscopy for velocity dispersion estimates.  It is not rare for $\lambda > 20$ clusters to appear close in projection and redshift to each other, and the overall density of clusters in this field is not particularly unusual.  However, three of these clusters have $\lambda>50$, and the spacing of two of these (2462, 2868) within $0.25$ deg and $\Delta z$ of $0.01$ is rare (2\% of $\lambda>50$ clusters in the redMaPPer catalog).

This superstructure, containing 4 of 21 outlier clusters, hints that a significant fraction of the outlier population originates from the presence of correlated structure.  Additional outlier clusters lie close in volume to each other and to other redMaPPer clusters, but again these associations are relatively common and the presence of nearby clusters alone is not sufficient to identify outliers.  The bottom panel of Figure~\ref{fig:superstructure}, shows the distribution in richness when the $\lambda$ calculation is scanned over redshift, $\lambda(z)$, for the outlier clusters in the volume-limited, redMaPPer catalog belonging to the $z\sim0.7$ superstructure and also including MEM\_MATCH\_ID 3610.  These distributions are compared to the $\lambda(z)$ expected for a cluster with no projection at $z=0.65$ and the 68\% and 95\% distributions of $\lambda(z)$ at the same redshift \citep{Costanzi19a}.  A wide $\lambda(z)$ may be an indication of significant line-of-sight structure. While a couple of the outlier clusters have $\lambda(z)$ that are somewhat wide compared to other clusters, particularly MEM\_MATCH\_ID 2868, they are generally within $\sim 2 \sigma$ of expectations for their redshifts.  In general, the full outlier population does not exhibit a significantly wider $\lambda(z)$ distribution compared to redMaPPer clusters at similar redshifts.  The fact that the outlier clusters are not clearly different in this metric highlights the difficulty of identifying complicated sight lines in photometric data.

\begin{figure}
    \centering
    \includegraphics[width = 1\columnwidth]{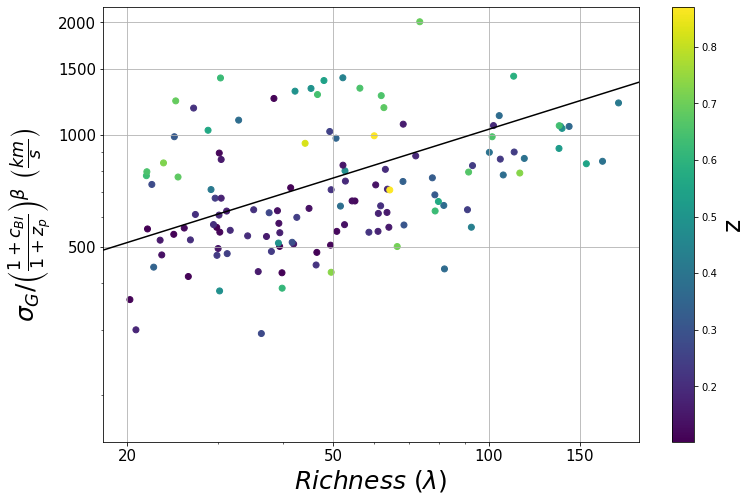}
    \caption{Velocity dispersion versus cluster richness with cluster redshift shown on the color axis. The outlier population appears to have a higher average redshift than the main population of clusters. The black line is the $\lambda$ adjusted $\sigma_v$--$\lambda$ relation found by \citet{redmapperIV}.}
    \label{fig:sigma_g_lambda_z}
\end{figure}

Projection effects in the redMaPPer catalog are expected to increase with redshift due to the fattening of the red-sequence and the difficulties associated with establishing photometric redshifts of high redshift galaxies.  Looking at the redshift distribution, the outlier population does appear to have a significantly higher average redshift than clusters with a similarly low richness. This is apparent in Figure \ref{fig:sigma_g_lambda_z} which shows $\sigma_G-\lambda$ color coded by redshift.  While overall the outliers make up 22\% of the $\lambda < 70$ clusters in our sample, they account for more than half of the $z>0.5$, $\lambda < 70$ clusters (11 out of 20). The presence of the outlier population and the redshift correlation is still present when limiting the sample to clusters in the volume limited redMaPPer cluster catalog with a redshift range of $z\in [0.2,0.65]$.

\begin{figure}
    \centering
    \includegraphics[width=0.5\textwidth]{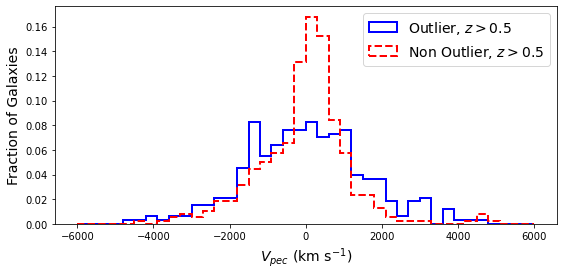}
    \caption{Histogram of the fractions of galaxies from the outlier (solid blue) and non-outlier (dashed red) populations at redshift $z>0.5$. At high redshift the non-outlier stacked population is narrower than the outlier stacked population. }
    \label{fig:outlier_z_stacked}
\end{figure}

Figure \ref{fig:outlier_z_stacked} shows the stacked histograms of the clusters with a redshift of $z>0.5$ which are outliers or non-outliers, respectively. The histogram of the outlier population is broader than that of the non-outlier population which suggests that the outlier population is not strictly due to the challenges associated with photometrically determining the redshift of red sequence galaxies at high redshifts. 

\begin{figure}
    \centering
    \includegraphics[width = 1\columnwidth]{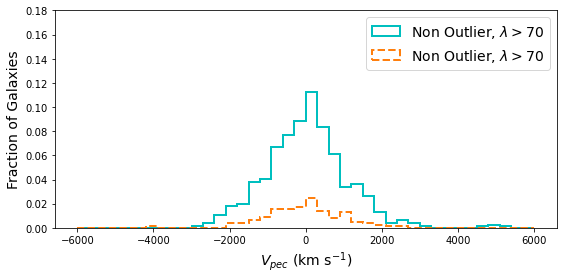}
    \includegraphics[width = 1\columnwidth]{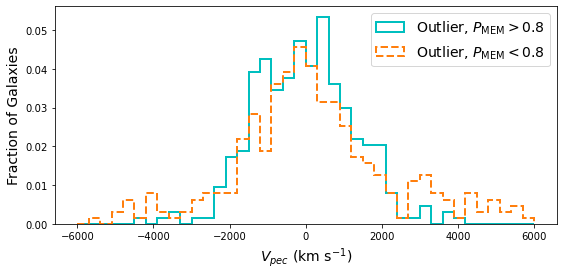}
    \includegraphics[width = 1\columnwidth]{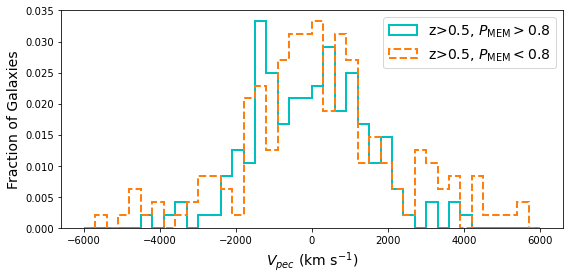}
    \includegraphics[width = 1\columnwidth]{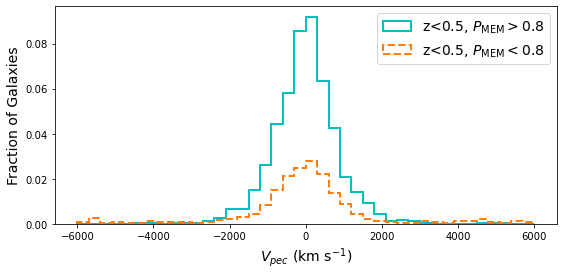}
    \caption{Stacked histograms of the fraction of galaxies from clusters in different populations in peculiar velocity bins separated by $P_{\rm MEM}$. Galaxies with $P_{\rm MEM}>0.8$ are shown by the solid blue histogram. Galaxies with $P_{\rm MEM}<0.8$ are shown by the dashed orange histogram. \textit{Top:} The main population at high richness which appears most similar to the outlier population in Figure \ref{fig:outlier_stacked}. \textit{Middle Top:} The outlier population. \textit{Middle Bottom:} Population of clusters with redshift $z>0.5$. \textit{Bottom:} Population of clusters with redshift $z<0.5$.}
    \label{fig:pmem_hist}
\end{figure}

An important question is whether the redMaPPer assigned richnesses of the outlier clusters correctly reflect their underlying mass.  $\lambda$ is computed as a sum of redMaPPer estimated galaxy membership probabilities, $P_{\rm MEM}$, with membership probability down weighted for galaxies as a function of distance in color and radius from the central cluster values.  While the redMaPPer $P_{\rm MEM}$ values are an indicator of whether a galaxy is more or less likely to be a cluster member and give $\lambda$'s which scale with mass with relatively low scatter, they are not a perfectly calibrated probability of cluster membership leading to biases in richness from projection and other effects \citep[][]{Rines18, Costanzi19a, Myles}. To explore the redMaPPer assigned richnesses of clusters in the outlier population, in Figure \ref{fig:pmem_hist} we investigate the velocity distributions as a function of $P_{\rm MEM}$. 

The top histogram shows the main cluster population at high richness separated for galaxies with $P_{\rm MEM}>0.8$ and $P_{\rm MEM}<0.8$. We chose a threshold of $P_{\rm MEM}=0.8$ as it provided similar results to that of $P_{\rm MEM}=0.5$ without drastically limiting our sample size. It is apparent that these clusters on average have few spectroscopic members with low $P_{\rm MEM}$. The second histogram from the top shows the outlier population. The low $P_{\rm MEM}$ galaxies form a broader distribution and account for a far more significant fraction of the galaxies in this population of clusters. This may be accounted for by the high average redshift of the outlier population as can be seen from the third histogram which shows the galaxies in clusters with redshift $z>0.5$. Again, the low $P_{\rm MEM}$ galaxies account for a large fraction of this population and have a slightly broader distribution. 

In general, redMaPPer clusters of similar richness at high redshift are composed of a larger number of potential cluster member galaxies with on average lower membership probabilities than their counterparts at lower redshift due to the increasing width of the red sequence and photometric redshift uncertainties.  In contrast the bottom histogram shows the stacked histogram of low-redshift, $z<0.5$ clusters. These clusters have a much smaller fraction of galaxies with low $P_{\rm MEM}$. The low $P_{\rm MEM}$ galaxies do form a broader distribution reflecting the fact that a larger fraction of these galaxies are not cluster members, as expected. 

The outlier clusters do not look substantially different than other high-redshift, redMaPPer clusters in terms of membership probabilities, while they do have wider velocity distributions (as seen in Figure \ref{fig:outlier_z_stacked}). We next turn to X-ray data where available to better understand the mass of these systems.

\subsection{Comparison to X-ray Properties}
\label{sec:5.3}

X-ray data where available can help distinguish massive from low mass clusters as well as allowing us to determine whether redMaPPer has chosen the correct central galaxy.  If the high velocity dispersions of the outlier clusters are indicative of a high mass, we expect to see luminous and hot X-ray emission.  In this case, the most likely reason for the low measured $\lambda$'s is miscentering by redMaPPer.  If instead the velocity dispersions are inflated by the projection of correlated structure, we would expect fainter or no X-ray emission.  The question in this case is whether the measured $\lambda$'s are consistent with the X-ray signal or if the richness calculation is also biased by projection effects.

\begin{figure}
    \centering
    \includegraphics[width = 1\columnwidth]{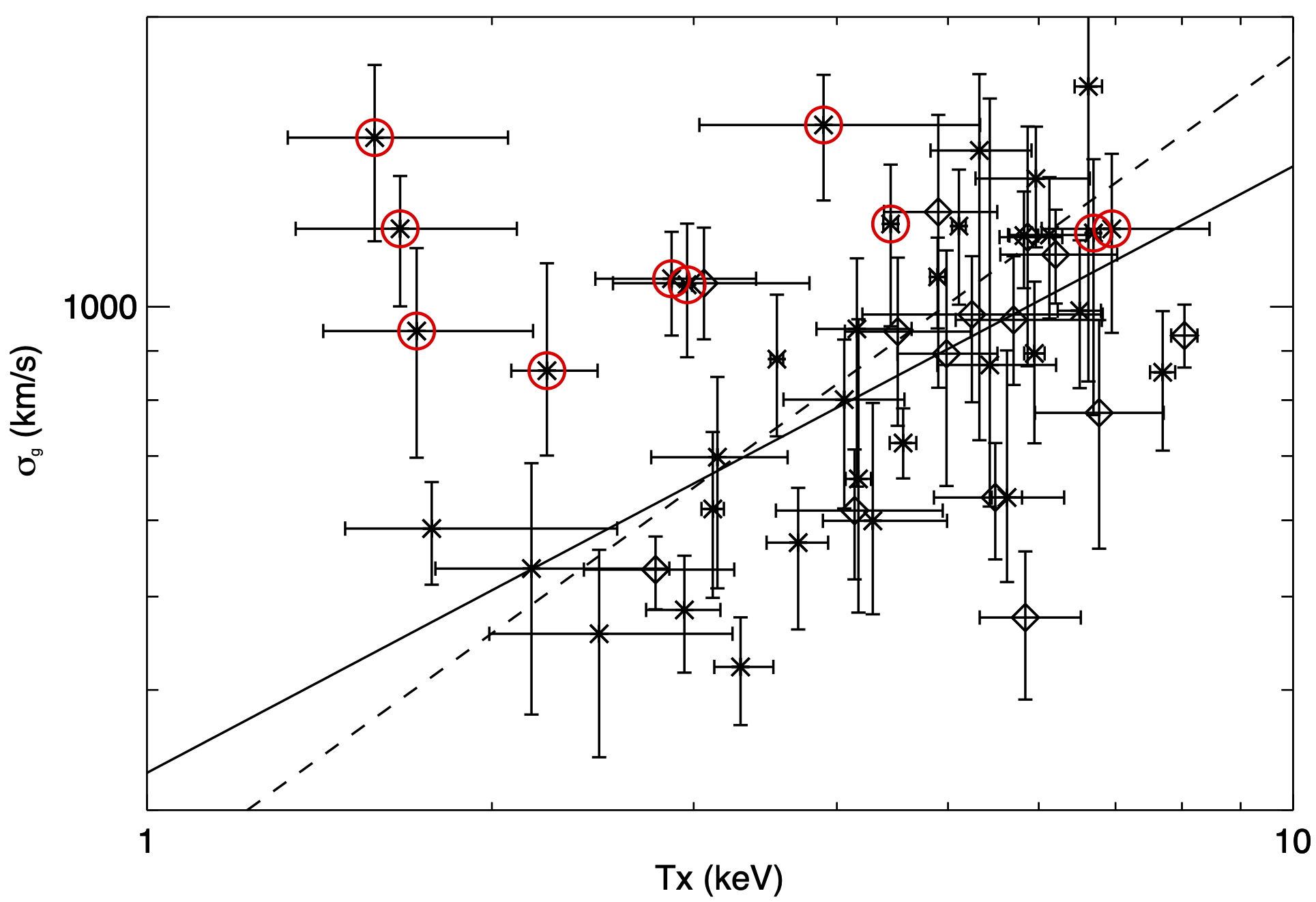}\\
    \vspace{0.2cm}
    \includegraphics[width = 1\columnwidth]{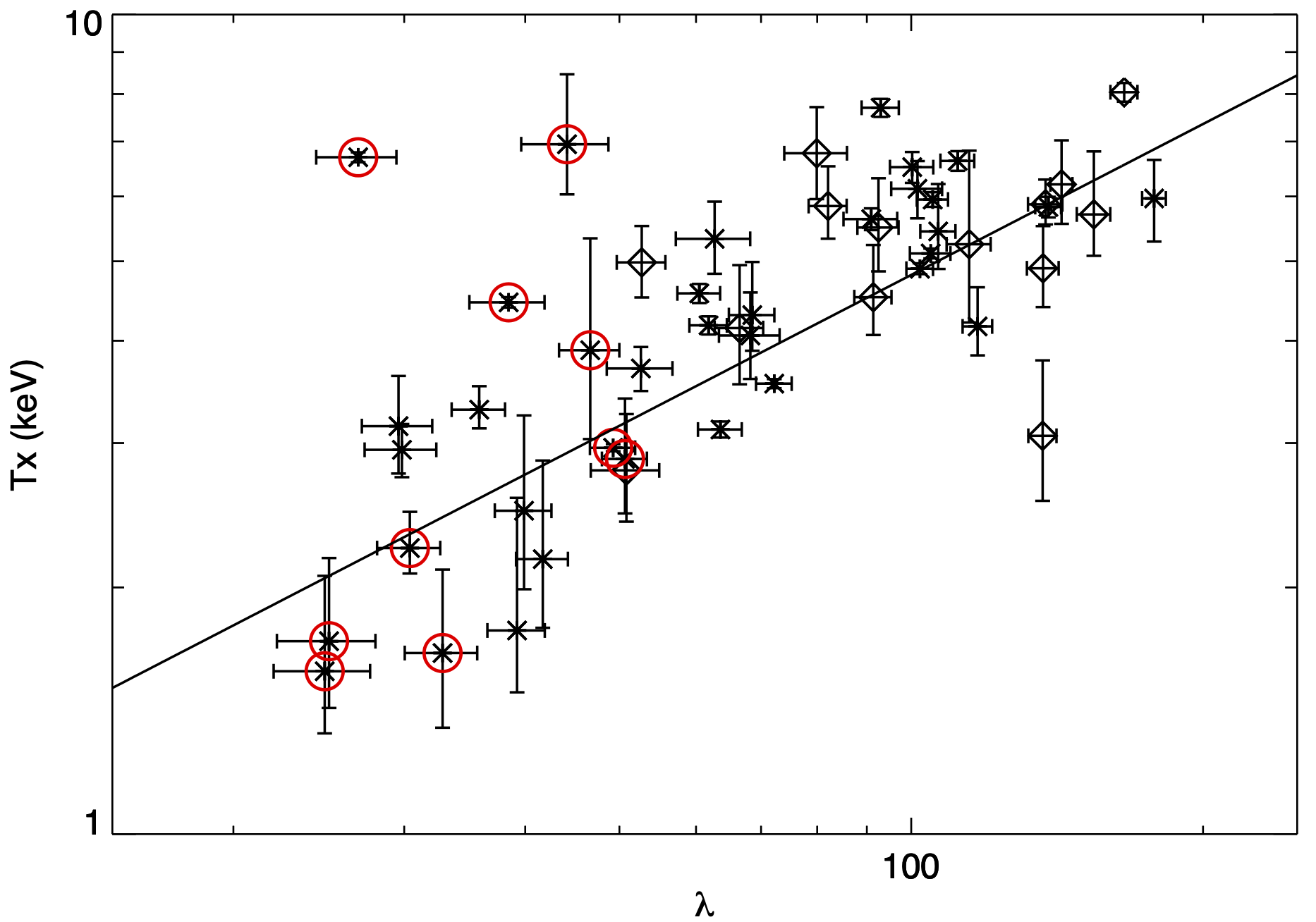}
    \caption{\textit{Top:} Velocity dispersion-temperature relation compared to the relations from \citet{FarahiXXL} (solid line) and \citet{Wilson16} (dashed line) for X-ray selected samples observed with XMM. \textit{Bottom:} Temperature-richness relation compared to the relation from \citet{Farahi19} for DES Y1 redMaPPer clusters (solid line).  In both plots, XMM temperature measurements are plotted with asterisks and Chandra measurements with diamonds.  Chandra temperatures have been adjusted to the XMM scale using the relation from \citet{redmapperSV}. The $T_X-\lambda$ relation from \citet{Farahi19} has likewise been adjusted to the XMM temperature scale. Velocity dispersion outlier clusters are circled in red.}
    \label{fig:txsig}
\end{figure}

\begin{figure}
    \centering
    \includegraphics[width = 1\columnwidth]{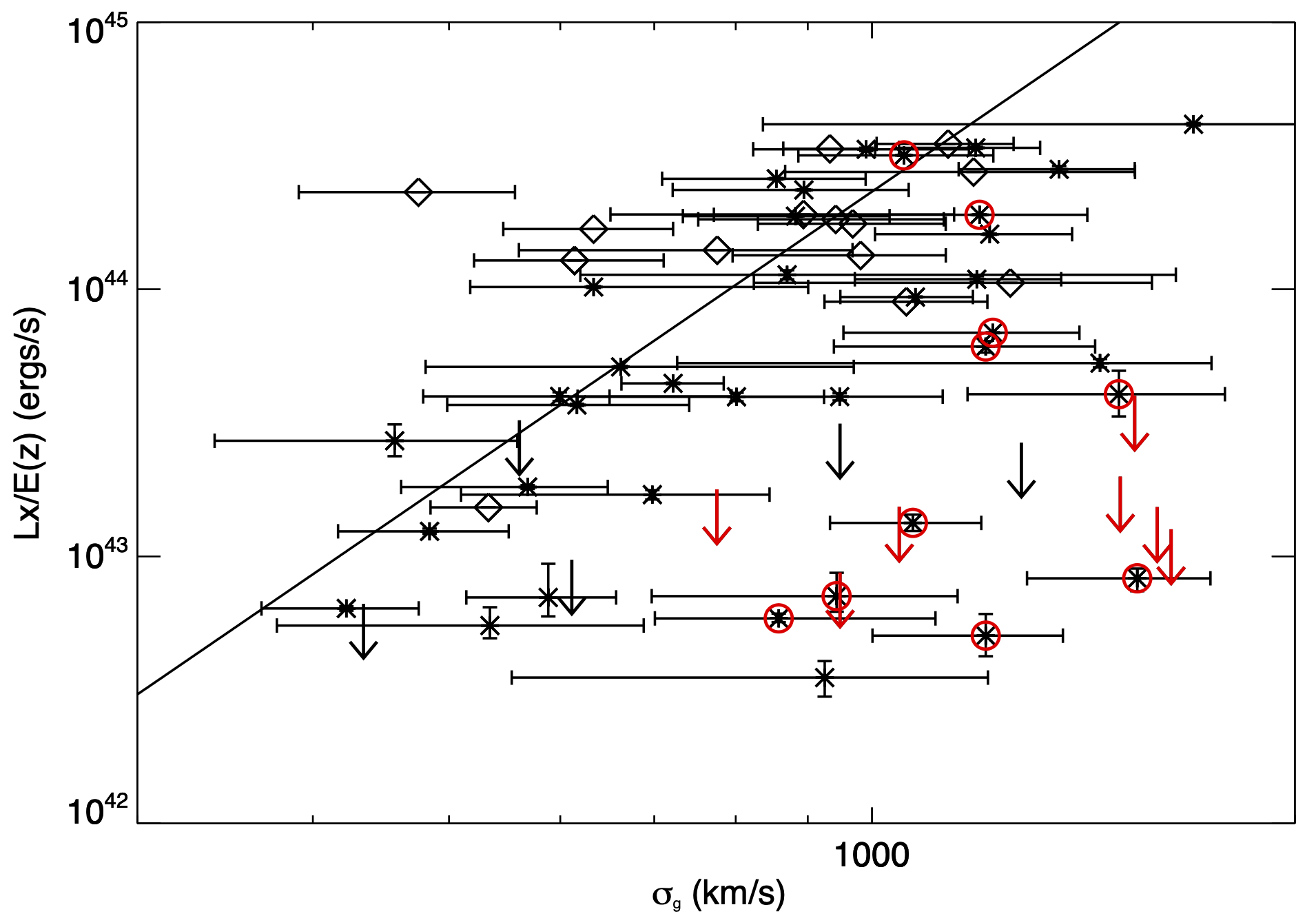}\\
    \vspace{0.2cm}
    \includegraphics[width = 1\columnwidth]{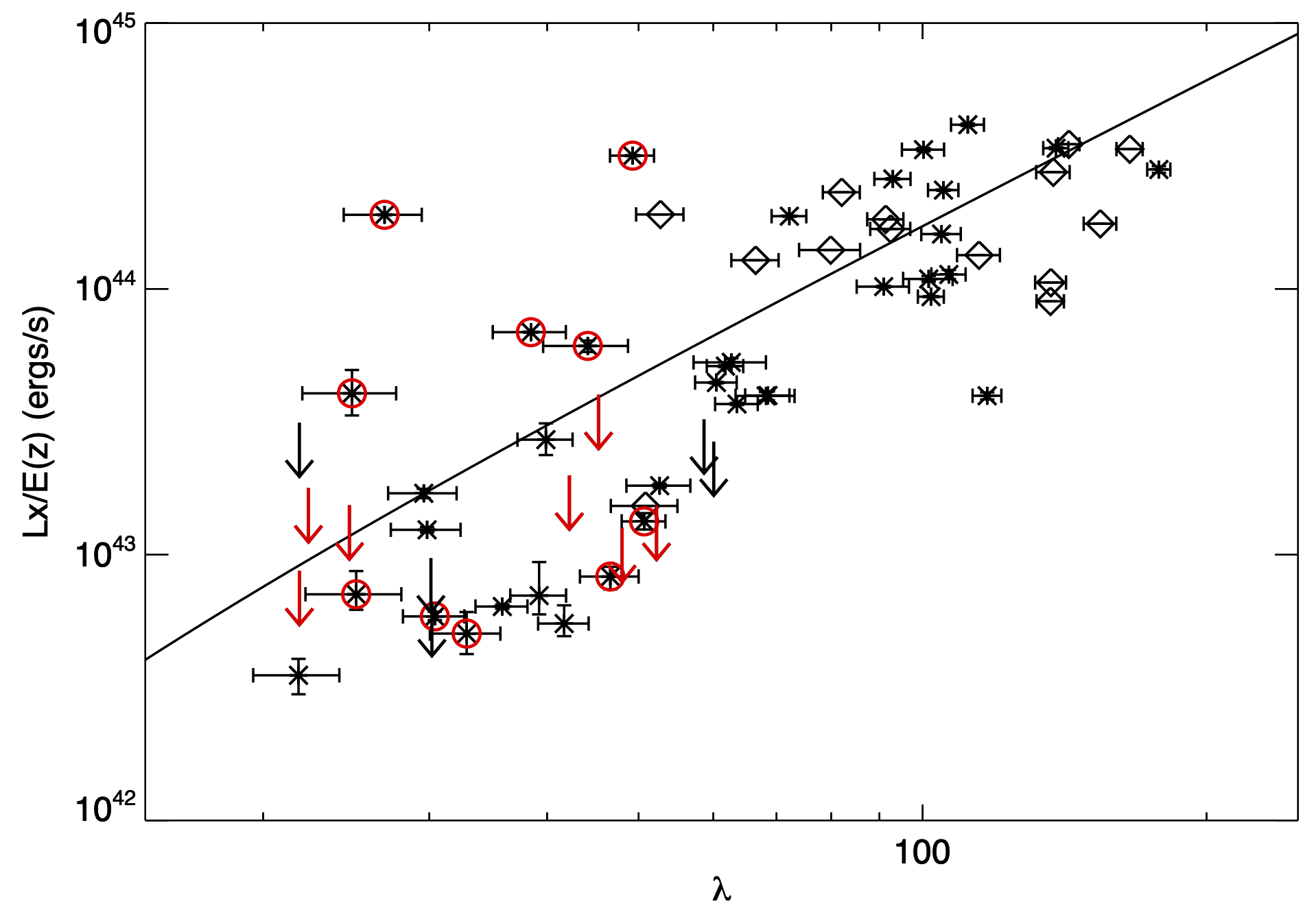}
    \caption{\textit{Top:} Luminosity-velocity dispersion relation compared to the relation from \citet{Popesso05} (solid line), specifically the relation from that reference with velocity dispersion determined from the red members, and we have converted the 0.1-2.4 keV luminosities to the 0.5-2 keV band. \textit{Bottom:} Luminosity-richness relation compared to the relation from \citet{matcha} (solid line) for SDSS redMaPPer clusters, specifically the relation from that reference which includes luminosity upper limits for undetected clusters.  In both plots, XMM measurements are plotted with asterisks and Chandra measurements with diamonds. Velocity dispersion outlier clusters are circled in red or plotted with red arrows.}
    \label{fig:lxsig}
\end{figure}

Figure \ref{fig:txsig} shows $\sigma_G-T_X$ and $T_X-\lambda$ for the clusters in our sample compared to relations from the literature, while Figure \ref{fig:lxsig} shows the $L_X-\sigma_G$ and $L_X-\lambda$ relations including upper limits for undetected clusters. There is a well known systematic offset between cluster X-ray temperatures estimated with XMM and Chandra \citep{Schell15}, and it is important when comparing the two to put them on the same scale.  We adjust the Chandra temperatures to the XMM scale using the relation in \cite{redmapperSV} derived through the comparison of 41 SDSS redMaPPer clusters observed with both instruments. Outlier clusters are circled in red.  

\begin{figure}
    \centering
    \includegraphics[width = 0.8\columnwidth]{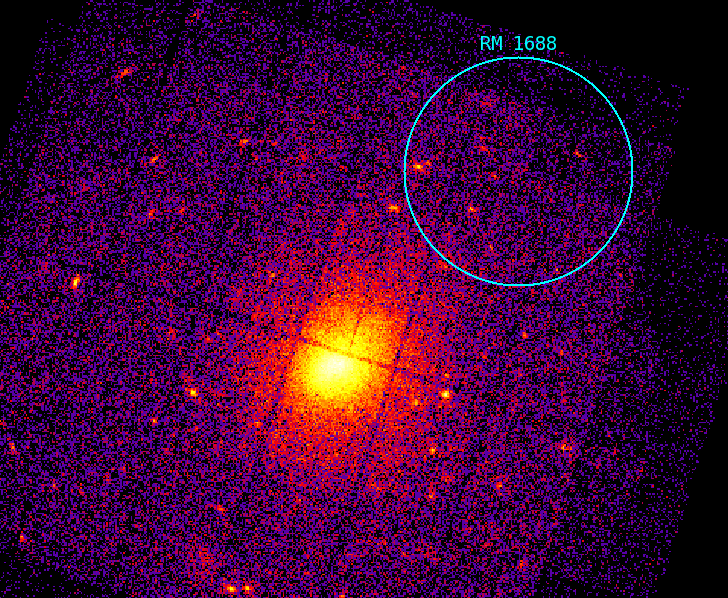}
    \includegraphics[width = 0.8\columnwidth]{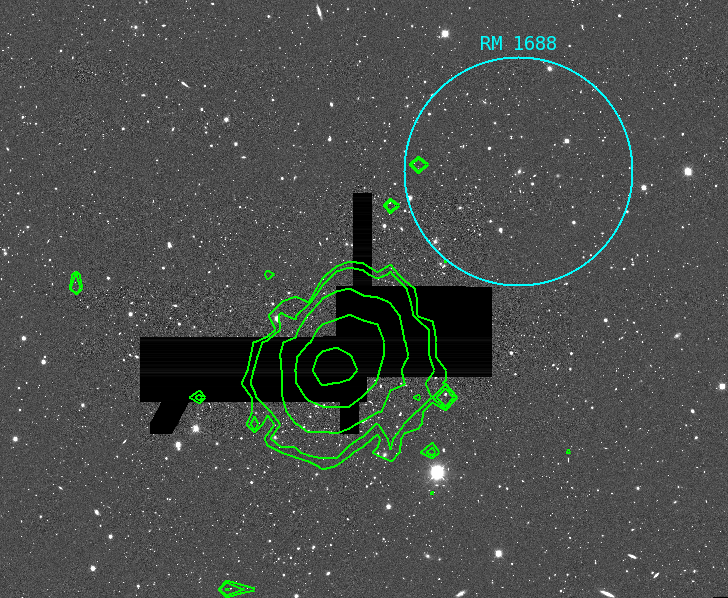}
    \caption{Example of a velocity disperion outlier, MEM\_MATCH\_ID  = 1688, that is X-ray bright, but miscentered by redMaPPer due to masking in the DES data.  The high velocity dispersion comes from sampling galaxies in the outskirts of a massive cluster. \textit{Top:} \textit{XMM-Newton} image of Abell 209 at $z=0.206$. RedMaPPer finds a low richness, $\lambda=27$ cluster with a similar redshift, $z=0.21$, offset from the X-ray cluster. The cyan circle marks the redMaPPer position and radius, $R_{\lambda} = 5$ arcmin. \textit{Bottom:} DES Y3A2 r-band image with X-ray contours overlaid in green and redMaPPer cluster region in cyan.  The position of Abell 209 is masked due to missing data.}
    \label{fig:miscenter}
\end{figure}

These figures reveal that the outliers in $\sigma_G-\lambda$ form a mixed population.  For some outlier clusters, the high veolcity dispersion is matched by a relatively high X-ray temperature.  In particular for two clusters, MEM\_MATCH\_ID 1688 with $T_X = 6.7$ keV and MEM\_MATCH\_ID 17296 with $T_X = 7.0$ keV, the high temperatures are inconsistent with the low measured richness.  The former of these clusters, 1688, is badly miscentered by redMaPPer, as shown in Figure~\ref{fig:miscenter}.  Missing DES data at the location of the X-ray bright cluster Abell 209 causes redMaPPer to miss the true center of this cluster; instead it finds a low richness system near the outskirts offset by 2.4 Mpc from the X-ray center. In general, another possibility would be that there is a separate group of galaxies near the massive, X-ray cluster; however, in this case, we have confirmed in the preliminary DES Y6 catalog, in which the missing DES data has been filled in, that redMaPPer finds a single, rich cluster at the location of the X-ray cluster.  The second, high $T_X$ cluster, 17296, has an estimated redshift of $z=0.82$ and is not in the volume-limited redMaPPer catalog. At these redshifts the richness estimate is less accurate as the depth is not sufficient to confindently detect fainter cluster galaxies.  Cluster 17296 is also miscentered, but only by 260 kpc with respect to the X-ray center, and recalculating the richness at the X-ray position does not significantly change the richness estimate.  Besides these two clusters, there are a couple of additional outlier clusters whose X-ray temperatures are somewhat high for their richnesses, but these are within the scatter in $T_X-\lambda$.  These same clusters are consistent within the scatter with the $\sigma_G-T_X$ relation.

The X-ray data indicate that miscentering is one reason for the velocity dispersion outliers, but not the dominant one.  Comparing to the X-ray peak position, we find that 6 of the 10 outlier clusters which are X-ray detected are miscentered by 100 kpc or more, a much higher fraction than for the cluster population overall.  However, with the exception of 1688 they are all miscentered by less than 1 Mpc, and their richness estimates increase by less than 15\% when centering on the X-ray position.  For 1688, the missing data means that we cannot calculate an appropriate richness, but this hot, Abell cluster would be expected to have a high richness.

A second component of the outlier clusters are detected in X-ray with lower temperatures and luminosities.  The velocity dispersions of these clusters are high compared to their X-ray properties and are likely inflated by correlated structure along the line of sight. A third portion of the outlier clusters are undetected in X-ray.  These non-detections are in most cases inconsistent with the high measured velocity dispersion, again pointing to contributions to the velocity dispersion of structure along the line of sight.

An interesting question is whether the richnesses of the outlier clusters reflect their X-ray properties or if they appear to be biased by projection and correlated structure.  Aside from 1688 and 17296 clusters, the richnesses of the X-ray detected clusters are consistent with their X-ray temperatures within the scatter. For most of the undetected clusters, the depth of the data is insufficient to judge, with the $L_X$ upper limits being consistent with richness at least within the large $L_X-\lambda$ scatter.  There is a tendency for the undetected, outlier clusters to scatter low in the $L_X-\lambda$ relation. While unclear from the current sample, this potentially indicates the richnesses of some of these clusters may be overestimated due to projection.  Deeper X-ray data is needed to confirm whether the undetected systems are truly virialized clusters or whether these are primarily projection effects where a filamentary structure or a string of small halos has been incorrectly identified as a significant cluster.

\section{Conclusion}
\label{sec:6}

In this paper, we calculate the velocity dispersions of galaxy clusters contained in the redMaPPer DES Y3 cluster catalog using available spectroscopic redshifts from external catalogs of galaxies identified as possible cluster member galaxies by redMaPPer.  Limiting the sample to clusters with sufficient statistics for velocity dispersion estimation, defined here as at least 15 spectroscopic members after interloper rejection, gives a total sample of 126 clusters.  The cluster velocity dispersions are examined as a function of richness, redshift, and X-ray properties.

Investigation of the velocity dispersions in comparison to cluster richness reveals a bimodal population. The main population follows a similar $\sigma_v-\lambda$ relation to that found by \citep{redmapperIV} for stacked spectroscopy of SDSS clusters.  However, there are a significant fraction of clusters with velocity disperisons that are high compared to their richnesses, referred to as the outlier population.  Defining outliers to be clusters whose lower limit on their velocity dispersion place them more than one standard deviation in the scatter high compared to the main population, this population makes up 17\% of the cluster sample and 22\% of clusters with richness $\lambda<70$.  These clusters tend to lie at higher redshifts, composing more than half (55\%) of $\lambda < 70$, $z>0.5$ clusters. However, they do have wider velocity distributions than non-outlier clusters at similar redshifts.  

Examination of the individual cluster velocity distributions and tests of a more conservative interloper rejection (Appendix \ref{sec:appendix}) indicate that the high velocity dispersions of the outliers do not appear to be the result of unrejected foreground or background galaxies or bimodal distributions in velocity space.  Most of these clusters simply appear to have wide, flat velocity distributions.  It remains possible, given our relatively sparse samples for some of these clusters, that a few of them have enhanced velocity dispersions due to the influence of unrejected interloping galaxies or mergers.  However, it is likely that many of these systems lie in regions with significant line-of-sight and correlated structure.  The photometric cluster selection, particularly at higher redshifts, can preferentially select this type of system \citep{Costanzi19a,DESKP, Wu}.  The outliers do not appear to have significantly different membership probability distributions or wider distributions of richness in redshift space, $\lambda(z)$, compared to clusters at similar redshifts, showing the difficulty in distinguishing line-of-sight structure from photometry.

Comparison to the cluster X-ray properties, where available, shows that a couple of the outlier clusters are hot, X-ray bright systems consistent with a high velocity dispersion and mass.  One of these clusters is a bright Abell cluster that is very miscentered by redMaPPer due to gaps in the DES data coverage at the cluster location.  However, most of the outlier clusters with X-ray data have low temperature and luminosity or are undetected in X-ray, implying lower mass systems.  Some of the outliers have richnesses consistent with their X-ray properties, but in general the sample size and depth of data in the case of non-detections is insufficient to make firm statements.  It is possible in some cases that the richnesses of these systems are overestimated due to projection effects.  The main cluster population has X-ray-$\sigma$ and X-ray-$\lambda$ relations similar to previous works.

In terms of the central velocity, we find that the standard deviation of the offset between the redMaPPer estimated cluster redshift from that of the biweight location calculated from the spectroscopy is $\sigma/(1+z)=0.0067$ which is similar to the previously established redMaPPer redshift scatter of ~0.006 \citep{mcclintock19}. The redMaPPer central galaxy offsets were found to have a small standard deviation of 0.0018; however there were several clusters with central galaxies that have velocity offsets up to 2000 \kms. These are likely the result of misidentification of the central by redMaPPer or cluster merging activity.  

Our results indicate that projection effects likely contribute significantly to redMaPPer cluster selection and possibly also richness estimation, particularly at lower richness and higher redshifts. In fact, modeling of the mass-richness relation using Sunyaev-Zel'dovich effect clusters from the South Pole Telescope SPT-SZ survey implies a growing contamination of redMaPPer samples as richness decreases by low mass objects boosted into the richness selected samples \citep{Grandis} with estimated fractions of contaminants consistent with the fraction of velocity dispersion outliers found here.  For lower redshift, SDSS redMaPPer clusters, \citet{Myles} also find that projection effects account for a growing fraction of the observed richness of lower richness clusters. A more quantitative understanding of these effects requires larger samples and more complete spectroscopy, particularly at high redshift, which is the goal of ongoing follow up efforts.

\section*{Acknowledgements}

This work was supported by the U.S. Department of Energy, Office of Science, Office of
High Energy Physics, under Award Numbers DE-SC0010107 and A00-1465-001.  AS is supported by the ERC-StG ‘ClustersXCosmo’ grant agreement 716762, by the FARE-MIUR grant 'ClustersXEuclid' R165SBKTMA, and by INFN InDark Grant. PTPV was supported by Fundação para a Ciência e a Tecnologia (FCT) through research grants UIDB/04434/2020 and UIDP/04434/2020.

Funding for the DES Projects has been provided by the U.S. Department of Energy, the U.S. National Science Foundation, the Ministry of Science and Education of Spain, 
the Science and Technology Facilities Council of the United Kingdom, the Higher Education Funding Council for England, the National Center for Supercomputing 
Applications at the University of Illinois at Urbana-Champaign, the Kavli Institute of Cosmological Physics at the University of Chicago, 
the Center for Cosmology and Astro-Particle Physics at the Ohio State University,
the Mitchell Institute for Fundamental Physics and Astronomy at Texas A\&M University, Financiadora de Estudos e Projetos, 
Funda{\c c}{\~a}o Carlos Chagas Filho de Amparo {\`a} Pesquisa do Estado do Rio de Janeiro, Conselho Nacional de Desenvolvimento Cient{\'i}fico e Tecnol{\'o}gico and 
the Minist{\'e}rio da Ci{\^e}ncia, Tecnologia e Inova{\c c}{\~a}o, the Deutsche Forschungsgemeinschaft and the Collaborating Institutions in the Dark Energy Survey. 

The Collaborating Institutions are Argonne National Laboratory, the University of California at Santa Cruz, the University of Cambridge, Centro de Investigaciones Energ{\'e}ticas, 
Medioambientales y Tecnol{\'o}gicas-Madrid, the University of Chicago, University College London, the DES-Brazil Consortium, the University of Edinburgh, 
the Eidgen{\"o}ssische Technische Hochschule (ETH) Z{\"u}rich, 
Fermi National Accelerator Laboratory, the University of Illinois at Urbana-Champaign, the Institut de Ci{\`e}ncies de l'Espai (IEEC/CSIC), 
the Institut de F{\'i}sica d'Altes Energies, Lawrence Berkeley National Laboratory, the Ludwig-Maximilians Universit{\"a}t M{\"u}nchen and the associated Excellence Cluster Universe, 
the University of Michigan, NFS's NOIRLab, the University of Nottingham, The Ohio State University, the University of Pennsylvania, the University of Portsmouth, 
SLAC National Accelerator Laboratory, Stanford University, the University of Sussex, Texas A\&M University, and the OzDES Membership Consortium.

Based in part on observations at Cerro Tololo Inter-American Observatory at NSF’s NOIRLab (NOIRLab Prop. ID 2012B-0001; PI: J. Frieman), which is managed by the Association of Universities for Research in Astronomy (AURA) under a cooperative agreement with the National Science Foundation.

The DES data management system is supported by the National Science Foundation under Grant Numbers AST-1138766 and AST-1536171.
The DES participants from Spanish institutions are partially supported by MICINN under grants ESP2017-89838, PGC2018-094773, PGC2018-102021, SEV-2016-0588, SEV-2016-0597, and MDM-2015-0509, some of which include ERDF funds from the European Union. IFAE is partially funded by the CERCA program of the Generalitat de Catalunya.
Research leading to these results has received funding from the European Research
Council under the European Union's Seventh Framework Program (FP7/2007-2013) including ERC grant agreements 240672, 291329, and 306478.
We  acknowledge support from the Brazilian Instituto Nacional de Ci\^encia
e Tecnologia (INCT) do e-Universo (CNPq grant 465376/2014-2).

This manuscript has been authored by Fermi Research Alliance, LLC under Contract No. DE-AC02-07CH11359 with the U.S. Department of Energy, Office of Science, Office of High Energy Physics.

\bibliographystyle{mnras}
\bibliography{references}

\begin{thebibliography}{}
\makeatletter
\relax
\def\mn@urlcharsother{\let\do\@makeother \do\$\do\&\do\#\do\^\do\_\do\%\do\~}
\def\mn@doi{\begingroup\mn@urlcharsother \@ifnextchar [ {\mn@doi@}
  {\mn@doi@[]}}
\def\mn@doi@[#1]#2{\def\@tempa{#1}\ifx\@tempa\@empty \href
  {http://dx.doi.org/#2} {doi:#2}\else \href {http://dx.doi.org/#2} {#1}\fi
  \endgroup}
\def\mn@eprint#1#2{\mn@eprint@#1:#2::\@nil}
\def\mn@eprint@arXiv#1{\href {http://arxiv.org/abs/#1} {{\tt arXiv:#1}}}
\def\mn@eprint@dblp#1{\href {http://dblp.uni-trier.de/rec/bibtex/#1.xml}
  {dblp:#1}}
\def\mn@eprint@#1:#2:#3:#4\@nil{\def\@tempa {#1}\def\@tempb {#2}\def\@tempc
  {#3}\ifx \@tempc \@empty \let \@tempc \@tempb \let \@tempb \@tempa \fi \ifx
  \@tempb \@empty \def\@tempb {arXiv}\fi \@ifundefined
  {mn@eprint@\@tempb}{\@tempb:\@tempc}{\expandafter \expandafter \csname
  mn@eprint@\@tempb\endcsname \expandafter{\@tempc}}}

\bibitem[\protect\citeauthoryear{{Abolfathi} et~al.,}{{Abolfathi}
  et~al.}{2018}]{SDSS}
{Abolfathi} B.,  et~al., 2018, \mn@doi [\apjs] {10.3847/1538-4365/aa9e8a},
  \href {https://ui.adsabs.harvard.edu/abs/2018ApJS..235...42A} {235, 42}

\bibitem[\protect\citeauthoryear{{Aguena} et~al.,}{{Aguena}
  et~al.}{2020}]{Aguena}
{Aguena} M.,  et~al., 2020, arXiv e-prints, \href
  {https://ui.adsabs.harvard.edu/abs/2020arXiv200808711A} {p. arXiv:2008.08711}

\bibitem[\protect\citeauthoryear{{Beers}, {Flynn}  \& {Gebhardt}}{{Beers}
  et~al.}{1990}]{Beers}
{Beers} T.~C.,  {Flynn} K.,   {Gebhardt} K.,  1990, \mn@doi [\aj]
  {10.1086/115487}, \href
  {https://ui.adsabs.harvard.edu/abs/1990AJ....100...32B} {100, 32}

\bibitem[\protect\citeauthoryear{{Bellagamba}, {Roncarelli}, {Maturi}  \&
  {Moscardini}}{{Bellagamba} et~al.}{2018}]{amico}
{Bellagamba} F.,  {Roncarelli} M.,  {Maturi} M.,   {Moscardini} L.,  2018,
  \mn@doi [\mnras] {10.1093/mnras/stx2701}, \href
  {https://ui.adsabs.harvard.edu/abs/2018MNRAS.473.5221B} {473, 5221}

\bibitem[\protect\citeauthoryear{{Bocquet} et~al.,}{{Bocquet}
  et~al.}{2019}]{spt2}
{Bocquet} S.,  et~al., 2019, \mn@doi [\apj] {10.3847/1538-4357/ab1f10}, \href
  {https://ui.adsabs.harvard.edu/abs/2019ApJ...878...55B} {878, 55}

\bibitem[\protect\citeauthoryear{{Childress} et~al.,}{{Childress}
  et~al.}{2017}]{ozdes}
{Childress} M.~J.,  et~al., 2017, \mn@doi [\mnras] {10.1093/mnras/stx1872},
  \href {https://ui.adsabs.harvard.edu/abs/2017MNRAS.472..273C} {472, 273}

\bibitem[\protect\citeauthoryear{{Costanzi} et~al.,}{{Costanzi}
  et~al.}{2019}]{Costanzi19a}
{Costanzi} M.,  et~al., 2019, \mn@doi [\mnras] {10.1093/mnras/sty2665}, \href
  {https://ui.adsabs.harvard.edu/abs/2019MNRAS.482..490C} {482, 490}

\bibitem[\protect\citeauthoryear{{DES Collaboration}}{{DES
  Collaboration}}{2005}]{DES2005}
{DES Collaboration} 2005, preprint, \href
  {http://adsabs.harvard.edu/abs/2005astro.ph.10346T} {} (\mn@eprint {arXiv}
  {astro-ph/0510346})

\bibitem[\protect\citeauthoryear{{DES Collaboration} et~al.,}{{DES
  Collaboration} et~al.}{2020}]{DESKP}
{DES Collaboration} et~al., 2020, arXiv e-prints, \href
  {https://ui.adsabs.harvard.edu/abs/2020arXiv200211124D} {p. arXiv:2002.11124}

\bibitem[\protect\citeauthoryear{{Dong}, {Pierpaoli}, {Gunn}  \&
  {Wechsler}}{{Dong} et~al.}{2008}]{Dong}
{Dong} F.,  {Pierpaoli} E.,  {Gunn} J.~E.,   {Wechsler} R.~H.,  2008, \mn@doi
  [\apj] {10.1086/522490}, \href
  {https://ui.adsabs.harvard.edu/abs/2008ApJ...676..868D} {676, 868}

\bibitem[\protect\citeauthoryear{{Durret} et~al.,}{{Durret}
  et~al.}{2011}]{Durret}
{Durret} F.,  et~al., 2011, \mn@doi [\aap] {10.1051/0004-6361/201116985}, \href
  {https://ui.adsabs.harvard.edu/abs/2011A&A...535A..65D} {535, A65}

\bibitem[\protect\citeauthoryear{{Farahi}, {Evrard}, {Rozo}, {Rykoff}  \&
  {Wechsler}}{{Farahi} et~al.}{2016}]{Farahi16}
{Farahi} A.,  {Evrard} A.~E.,  {Rozo} E.,  {Rykoff} E.~S.,   {Wechsler} R.~H.,
  2016, \mn@doi [\mnras] {10.1093/mnras/stw1143}, \href
  {https://ui.adsabs.harvard.edu/abs/2016MNRAS.460.3900F} {460, 3900}

\bibitem[\protect\citeauthoryear{{Farahi} et~al.,}{{Farahi}
  et~al.}{2018}]{FarahiXXL}
{Farahi} A.,  et~al., 2018, \mn@doi [\aap] {10.1051/0004-6361/201731321}, \href
  {https://ui.adsabs.harvard.edu/abs/2018A&A...620A...8F} {620, A8}

\bibitem[\protect\citeauthoryear{{Farahi} et~al.,}{{Farahi}
  et~al.}{2019}]{Farahi19}
{Farahi} A.,  et~al., 2019, \mn@doi [\mnras] {10.1093/mnras/stz2689}, \href
  {https://ui.adsabs.harvard.edu/abs/2019MNRAS.490.3341F} {490, 3341}

\bibitem[\protect\citeauthoryear{{Ferragamo}, {Rubi{\~n}o-Mart{\'\i}n},
  {Betancort-Rijo}, {Munari}, {Sartoris}  \& {Barrena}}{{Ferragamo}
  et~al.}{2020}]{Ferragamo}
{Ferragamo} A.,  {Rubi{\~n}o-Mart{\'\i}n} J.~A.,  {Betancort-Rijo} J.,
  {Munari} E.,  {Sartoris} B.,   {Barrena} R.,  2020, \mn@doi [\aap]
  {10.1051/0004-6361/201834837}, \href
  {https://ui.adsabs.harvard.edu/abs/2020A&A...641A..41F} {641, A41}

\bibitem[\protect\citeauthoryear{{Giles} et~al.,}{{Giles} et~al.}{2022}]{Giles}
{Giles} P.~A.,  et~al., 2022, arXiv e-prints, \href
  {https://ui.adsabs.harvard.edu/abs/2022arXiv220211107G} {p. arXiv:2202.11107}

\bibitem[\protect\citeauthoryear{{Gladders} \& {Yee}}{{Gladders} \&
  {Yee}}{2005}]{Gladders}
{Gladders} M.~D.,  {Yee} H.~K.~C.,  2005, \mn@doi [\apjs] {10.1086/427327},
  \href {https://ui.adsabs.harvard.edu/abs/2005ApJS..157....1G} {157, 1}

\bibitem[\protect\citeauthoryear{{Grandis} et~al.,}{{Grandis}
  et~al.}{2021}]{Grandis}
{Grandis} S.,  et~al., 2021, \mn@doi [\mnras] {10.1093/mnras/stab869}, \href
  {https://ui.adsabs.harvard.edu/abs/2021MNRAS.504.1253G} {504, 1253}

\bibitem[\protect\citeauthoryear{{Gschwend} et~al.,}{{Gschwend}
  et~al.}{2018}]{portal}
{Gschwend} J.,  et~al., 2018, \mn@doi [Astronomy and Computing]
  {10.1016/j.ascom.2018.08.008}, \href
  {https://ui.adsabs.harvard.edu/abs/2018A&C....25...58G} {25, 58}

\bibitem[\protect\citeauthoryear{{Hollowood} et~al.,}{{Hollowood}
  et~al.}{2019}]{matcha}
{Hollowood} D.~L.,  et~al., 2019, \mn@doi [\apjs] {10.3847/1538-4365/ab3d27},
  \href {https://ui.adsabs.harvard.edu/abs/2019ApJS..244...22H} {244, 22}

\bibitem[\protect\citeauthoryear{{Kim}, {Peter}  \& {Wittman}}{{Kim}
  et~al.}{2017}]{Kim17}
{Kim} S.~Y.,  {Peter} A. H.~G.,   {Wittman} D.,  2017, \mn@doi [\mnras]
  {10.1093/mnras/stx896}, \href
  {https://ui.adsabs.harvard.edu/abs/2017MNRAS.469.1414K} {469, 1414}

\bibitem[\protect\citeauthoryear{{Koester} et~al.,}{{Koester}
  et~al.}{2007}]{maxBCG}
{Koester} B.~P.,  et~al., 2007, \mn@doi [\apj] {10.1086/509599}, \href
  {https://ui.adsabs.harvard.edu/abs/2007ApJ...660..239K} {660, 239}

\bibitem[\protect\citeauthoryear{{Licitra}, {Mei}, {Raichoor}, {Erben}  \&
  {Hildebrandt}}{{Licitra} et~al.}{2016}]{Licitra}
{Licitra} R.,  {Mei} S.,  {Raichoor} A.,  {Erben} T.,   {Hildebrandt} H.,
  2016, \mn@doi [\mnras] {10.1093/mnras/stv2309}, \href
  {https://ui.adsabs.harvard.edu/abs/2016MNRAS.455.3020L} {455, 3020}

\bibitem[\protect\citeauthoryear{{Lidman} et~al.,}{{Lidman}
  et~al.}{2020}]{Lidman}
{Lidman} C.,  et~al., 2020, \mn@doi [\mnras] {10.1093/mnras/staa1341}, \href
  {https://ui.adsabs.harvard.edu/abs/2020MNRAS.496...19L} {496, 19}

\bibitem[\protect\citeauthoryear{{Lloyd-Davies} et~al.,}{{Lloyd-Davies}
  et~al.}{2011}]{XCS}
{Lloyd-Davies} E.~J.,  et~al., 2011, \mn@doi [\mnras]
  {10.1111/j.1365-2966.2011.19117.x}, \href
  {http://adsabs.harvard.edu/abs/2011MNRAS.418...14L} {418, 14}

\bibitem[\protect\citeauthoryear{{Lucey}}{{Lucey}}{1983}]{Lucey83}
{Lucey} J.~R.,  1983, \mn@doi [\mnras] {10.1093/mnras/204.1.33}, \href
  {https://ui.adsabs.harvard.edu/abs/1983MNRAS.204...33L} {204, 33}

\bibitem[\protect\citeauthoryear{{Mantz}, {Allen}, {Rapetti}  \&
  {Ebeling}}{{Mantz} et~al.}{2010}]{Mantz10}
{Mantz} A.,  {Allen} S.~W.,  {Rapetti} D.,   {Ebeling} H.,  2010, \mn@doi
  [\mnras] {10.1111/j.1365-2966.2010.16992.x}, \href
  {https://ui.adsabs.harvard.edu/abs/2010MNRAS.406.1759M} {406, 1759}

\bibitem[\protect\citeauthoryear{{Mantz} et~al.,}{{Mantz}
  et~al.}{2015}]{Mantz15}
{Mantz} A.~B.,  et~al., 2015, \mn@doi [\mnras] {10.1093/mnras/stu2096}, \href
  {https://ui.adsabs.harvard.edu/abs/2015MNRAS.446.2205M} {446, 2205}

\bibitem[\protect\citeauthoryear{{McClintock} et~al.,}{{McClintock}
  et~al.}{2019}]{mcclintock19}
{McClintock} T.,  et~al., 2019, \mn@doi [\mnras] {10.1093/mnras/sty2711}, \href
  {https://ui.adsabs.harvard.edu/abs/2019MNRAS.482.1352M} {482, 1352}

\bibitem[\protect\citeauthoryear{{Milkeraitis}, {van Waerbeke}, {Heymans},
  {Hildebrand t}, {Dietrich}  \& {Erben}}{{Milkeraitis}
  et~al.}{2010}]{Milkeraitis}
{Milkeraitis} M.,  {van Waerbeke} L.,  {Heymans} C.,  {Hildebrand t} H.,
  {Dietrich} J.~P.,   {Erben} T.,  2010, \mn@doi [\mnras]
  {10.1111/j.1365-2966.2010.16720.x}, \href
  {https://ui.adsabs.harvard.edu/abs/2010MNRAS.406..673M} {406, 673}

\bibitem[\protect\citeauthoryear{{Murphy}, {Geach}  \& {Bower}}{{Murphy}
  et~al.}{2012}]{Murphy}
{Murphy} D.~N.~A.,  {Geach} J.~E.,   {Bower} R.~G.,  2012, \mn@doi [\mnras]
  {10.1111/j.1365-2966.2011.19782.x}, \href
  {https://ui.adsabs.harvard.edu/abs/2012MNRAS.420.1861M} {420, 1861}

\bibitem[\protect\citeauthoryear{{Myles} et~al.,}{{Myles} et~al.}{2020}]{Myles}
{Myles} J.~T.,  et~al., 2020, arXiv e-prints, \href
  {https://ui.adsabs.harvard.edu/abs/2020arXiv201107070M} {p. arXiv:2011.07070}

\bibitem[\protect\citeauthoryear{{Oguri}}{{Oguri}}{2014}]{Oguri}
{Oguri} M.,  2014, \mn@doi [\mnras] {10.1093/mnras/stu1446}, \href
  {https://ui.adsabs.harvard.edu/abs/2014MNRAS.444..147O} {444, 147}

\bibitem[\protect\citeauthoryear{{Planck Collaboration} et~al.,}{{Planck
  Collaboration} et~al.}{2016}]{planck1}
{Planck Collaboration} et~al., 2016, \mn@doi [\aap]
  {10.1051/0004-6361/201525833}, \href
  {https://ui.adsabs.harvard.edu/abs/2016A&A...594A..24P} {594, A24}

\bibitem[\protect\citeauthoryear{{Popesso}, {Biviano}, {B{\"o}hringer},
  {Romaniello}  \& {Voges}}{{Popesso} et~al.}{2005}]{Popesso05}
{Popesso} P.,  {Biviano} A.,  {B{\"o}hringer} H.,  {Romaniello} M.,   {Voges}
  W.,  2005, \mn@doi [\aap] {10.1051/0004-6361:20041915}, \href
  {https://ui.adsabs.harvard.edu/abs/2005A&A...433..431P} {433, 431}

\bibitem[\protect\citeauthoryear{{Rines}, {Geller}, {Diaferio}, {Hwang}  \&
  {Sohn}}{{Rines} et~al.}{2018}]{Rines18}
{Rines} K.~J.,  {Geller} M.~J.,  {Diaferio} A.,  {Hwang} H.~S.,   {Sohn} J.,
  2018, \mn@doi [\apj] {10.3847/1538-4357/aacd49}, \href
  {https://ui.adsabs.harvard.edu/abs/2018ApJ...862..172R} {862, 172}

\bibitem[\protect\citeauthoryear{{Rozo} et~al.,}{{Rozo} et~al.}{2010}]{Rozo10}
{Rozo} E.,  et~al., 2010, \mn@doi [\apj] {10.1088/0004-637X/708/1/645}, \href
  {https://ui.adsabs.harvard.edu/abs/2010ApJ...708..645R} {708, 645}

\bibitem[\protect\citeauthoryear{{Rozo}, {Rykoff}, {Becker}, {Reddick}  \&
  {Wechsler}}{{Rozo} et~al.}{2015}]{redmapperIV}
{Rozo} E.,  {Rykoff} E.~S.,  {Becker} M.,  {Reddick} R.~M.,   {Wechsler} R.~H.,
   2015, \mn@doi [\mnras] {10.1093/mnras/stv1560}, \href
  {http://adsabs.harvard.edu/abs/2015MNRAS.453...38R} {453, 38}

\bibitem[\protect\citeauthoryear{{Ruel} et~al.,}{{Ruel}
  et~al.}{2014}]{ruel_optical_2014}
{Ruel} J.,  et~al., 2014, \mn@doi [\apj] {10.1088/0004-637X/792/1/45}, \href
  {https://ui.adsabs.harvard.edu/abs/2014ApJ...792...45R} {792, 45}

\bibitem[\protect\citeauthoryear{{Rykoff} et~al.,}{{Rykoff}
  et~al.}{2014}]{redmapperI}
{Rykoff} E.~S.,  et~al., 2014, \mn@doi [\apj] {10.1088/0004-637X/785/2/104},
  \href {http://adsabs.harvard.edu/abs/2014ApJ...785..104R} {785, 104}

\bibitem[\protect\citeauthoryear{{Rykoff} et~al.,}{{Rykoff}
  et~al.}{2016}]{redmapperSV}
{Rykoff} E.~S.,  et~al., 2016, \mn@doi [\apjs] {10.3847/0067-0049/224/1/1},
  \href {http://adsabs.harvard.edu/abs/2016ApJS..224....1R} {224, 1}

\bibitem[\protect\citeauthoryear{{Saro}, {Mohr}, {Bazin}  \& {Dolag}}{{Saro}
  et~al.}{2013}]{Saro13}
{Saro} A.,  {Mohr} J.~J.,  {Bazin} G.,   {Dolag} K.,  2013, \mn@doi [\apj]
  {10.1088/0004-637X/772/1/47}, \href
  {https://ui.adsabs.harvard.edu/abs/2013ApJ...772...47S} {772, 47}

\bibitem[\protect\citeauthoryear{{Schellenberger}, {Reiprich}, {Lovisari},
  {Nevalainen}  \& {David}}{{Schellenberger} et~al.}{2015}]{Schell15}
{Schellenberger} G.,  {Reiprich} T.~H.,  {Lovisari} L.,  {Nevalainen} J.,
  {David} L.,  2015, \mn@doi [\aap] {10.1051/0004-6361/201424085}, \href
  {https://ui.adsabs.harvard.edu/abs/2015A&A...575A..30S} {575, A30}

\bibitem[\protect\citeauthoryear{{Sevilla-Noarbe} et~al.,}{{Sevilla-Noarbe}
  et~al.}{2020}]{Gold}
{Sevilla-Noarbe} I.,  et~al., 2020, arXiv e-prints, \href
  {https://ui.adsabs.harvard.edu/abs/2020arXiv201103407S} {p. arXiv:2011.03407}

\bibitem[\protect\citeauthoryear{{Soares-Santos} et~al.,}{{Soares-Santos}
  et~al.}{2011}]{Soares-Santos}
{Soares-Santos} M.,  et~al., 2011, \mn@doi [\apj] {10.1088/0004-637X/727/1/45},
  \href {https://ui.adsabs.harvard.edu/abs/2011ApJ...727...45S} {727, 45}

\bibitem[\protect\citeauthoryear{{Sohn}, {Geller}, {Rines}, {Hwang}, {Utsumi}
  \& {Diaferio}}{{Sohn} et~al.}{2018}]{Sohn}
{Sohn} J.,  {Geller} M.~J.,  {Rines} K.~J.,  {Hwang} H.~S.,  {Utsumi} Y.,
  {Diaferio} A.,  2018, \mn@doi [\apj] {10.3847/1538-4357/aab20b}, \href
  {https://ui.adsabs.harvard.edu/abs/2018ApJ...856..172S} {856, 172}

\bibitem[\protect\citeauthoryear{{Sunayama} et~al.,}{{Sunayama}
  et~al.}{2020}]{Sunayama}
{Sunayama} T.,  et~al., 2020, \mn@doi [\mnras] {10.1093/mnras/staa1646}, \href
  {https://ui.adsabs.harvard.edu/abs/2020MNRAS.496.4468S} {496, 4468}

\bibitem[\protect\citeauthoryear{{Vikhlinin} et~al.,}{{Vikhlinin}
  et~al.}{2009}]{Vikhlinin09}
{Vikhlinin} A.,  et~al., 2009, \mn@doi [\apj] {10.1088/0004-637X/692/2/1060},
  \href {https://ui.adsabs.harvard.edu/abs/2009ApJ...692.1060V} {692, 1060}

\bibitem[\protect\citeauthoryear{{Weinberg}, {Mortonson}, {Eisenstein},
  {Hirata}, {Riess}  \& {Rozo}}{{Weinberg} et~al.}{2013}]{Weinberg13}
{Weinberg} D.~H.,  {Mortonson} M.~J.,  {Eisenstein} D.~J.,  {Hirata} C.,
  {Riess} A.~G.,   {Rozo} E.,  2013, \mn@doi [\physrep]
  {10.1016/j.physrep.2013.05.001}, \href
  {https://ui.adsabs.harvard.edu/abs/2013PhR...530...87W} {530, 87}

\bibitem[\protect\citeauthoryear{{Wilson} et~al.,}{{Wilson}
  et~al.}{2016}]{Wilson16}
{Wilson} S.,  et~al., 2016, \mn@doi [\mnras] {10.1093/mnras/stw1947}, \href
  {https://ui.adsabs.harvard.edu/abs/2016MNRAS.463..413W} {463, 413}

\bibitem[\protect\citeauthoryear{{Wu} et~al.,}{{Wu} et~al.}{2022}]{Wu}
{Wu} H.-Y.,  et~al., 2022, arXiv e-prints, \href
  {https://ui.adsabs.harvard.edu/abs/2022arXiv220305416W} {p. arXiv:2203.05416}

\bibitem[\protect\citeauthoryear{{Zhang} et~al.,}{{Zhang}
  et~al.}{2019}]{Zhang19}
{Zhang} Y.,  et~al., 2019, \mn@doi [\mnras] {10.1093/mnras/stz1361}, \href
  {https://ui.adsabs.harvard.edu/abs/2019MNRAS.487.2578Z} {487, 2578}

\bibitem[\protect\citeauthoryear{{de Haan} et~al.,}{{de Haan}
  et~al.}{2016}]{spt1}
{de Haan} T.,  et~al., 2016, \mn@doi [\apj] {10.3847/0004-637X/832/1/95}, \href
  {https://ui.adsabs.harvard.edu/abs/2016ApJ...832...95D} {832, 95}

\makeatother
\end{thebibliography}

\section{DATA AVAILABILITY}
The data underlying this article are available at https://des.ncsa.illinois.edu/releases/dr2/dr2-docs and https://docs.datacentral.org.au/ozdes/overview/ozdes-data-release/.  The redMaPPer catalog used  proprietary to the Dark Energy Survey Collaboration, but will be released upon publication of the Y3 cluster cosmology papers.

\appendix
\section{Velocity Distributions and Interloper Rejection}
\label{sec:appendix}

\begin{figure}
    \centering
    \includegraphics[width=0.5\textwidth]{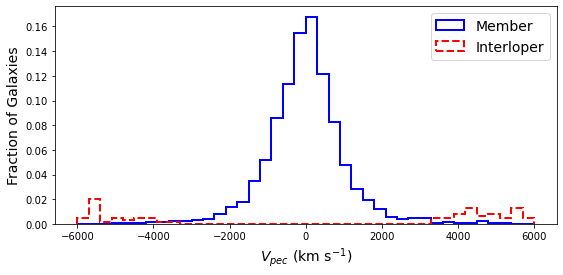}
    \caption{Histogram of the fraction of all clusters stacked together in peculiar velocity bins. The solid blue line represents galaxies included in the velocity dispersion calculations, the dashed red line shows the interloping galaxies. A large fraction of the interloping galaxies have absolute peculiar velocities larger than 6000 \kms and thus are not shown.}
    \label{fig:all_stacked}
\end{figure}

In this Appendix, we present peculiar velocity histograms with their corresponding bootstrap $\sigma_G$ distribution for all clusters in our sample (Figure \ref{fig:galery}) and the stacked histogram for all clusters (Figure \ref{fig:all_stacked}).  We also explore the effect of making a cut on $P_{\rm MEM}$ or a more conservative initial interloper rejection.  

The initial removal of interlopers was performed using the richness dependent cut from section \ref{sec:3}. Interlopers were further rejected by a 3$\sigma$ cut applied and iterated on for both the biweight and gapper methods. The interlopers found using this method are shown in red in Figure \ref{fig:galery}. There are a number interloping galaxies with peculiar velocity differences greater than 4000 \kms which are not shown in Figure \ref{fig:galery}. To examine our interloper rejection we stacked all of the clusters together in Figure \ref{fig:all_stacked}. This figure shows an overall good separation of interloper galaxies from the central cluster component, though the accuracy for individual clusters will vary given the spectroscopic sampling. There are several clusters with non-rejected members that appear to have large velocity offsets from the main galaxy population (151, 205, 6483, etc.). While these members would skew a single velocity dispersion statistic of the cluster, bootstrapping provides for a more robust velocity dispersion estimate with confidence intervals that accurately represent the probability distribution of $\sigma_G$.

\begin{figure*}
    \caption{Gallery of all clusters studied with corresponding $\sigma_G$ bootstrap distributions. The cluster MEM\_MATCH\_ID is listed in the top left of each subplot, outlier clusters are denoted with a red colored ID as well as an asterisk. For the peculiar velocity ($v_{pec}$) plots member galaxies are shown in blue, interloping galaxies are shown in red. For the $\sigma_G$ distributions the black line shows our reported $\sigma_G$ for that cluster and the grey bar covers the $\sigma_G$ confidence interval for that cluster.}
    \label{fig:galery}
    \begin{subfigure}{\textwidth}
        \centering
        \includegraphics[height=0.9\textheight]{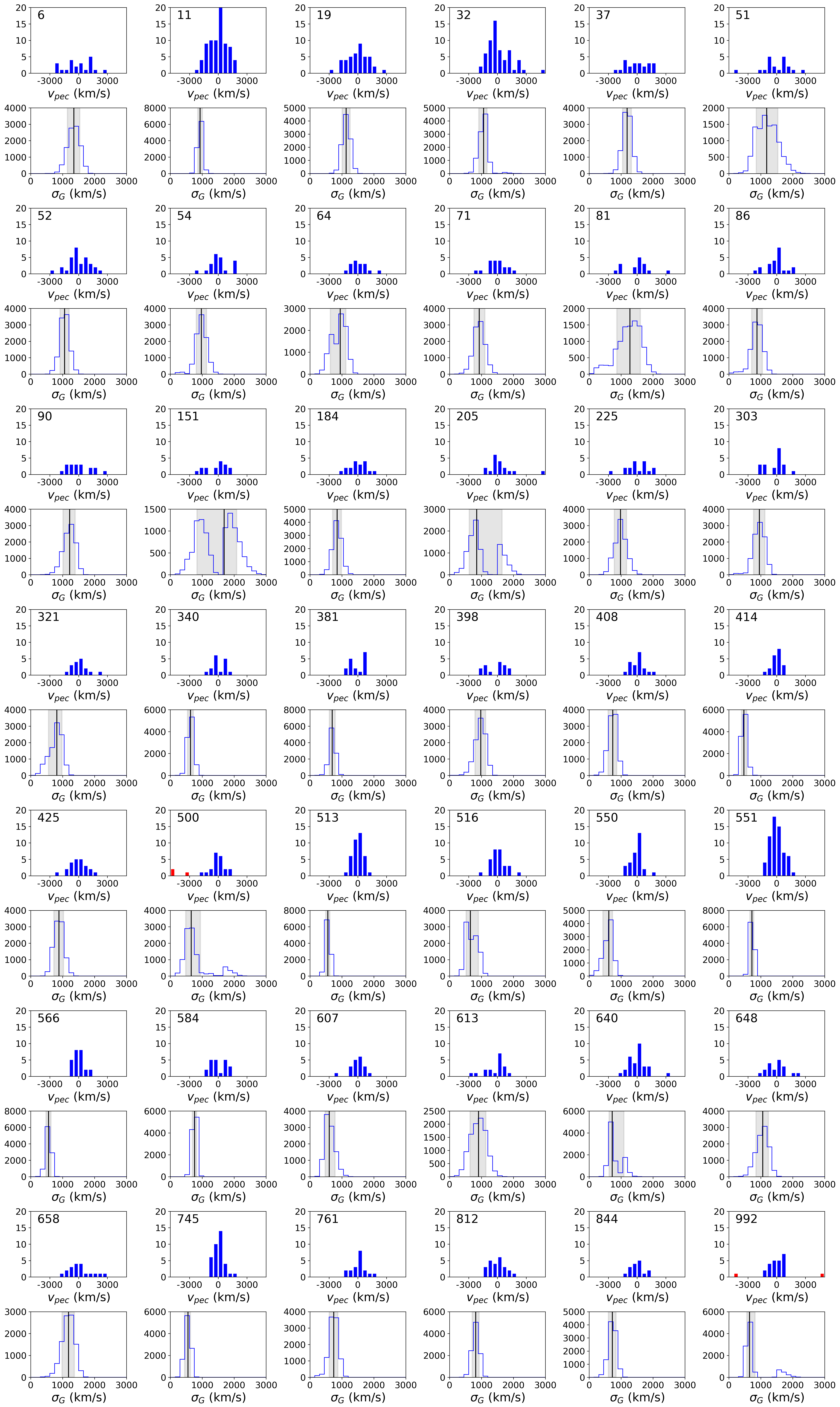}
        \caption{Gallery of all studied clusters. Cluster redMaPPer ID is shown in top left of each plot. Member galaxies are shown in blue, interlopers are shown in red. Outlier clusters are denoted with a red colored ID as well as an asterisk.}
    \end{subfigure}
\end{figure*}
\begin{figure*}
    \ContinuedFloat
    \begin{subfigure}{\textwidth}
        \centering
        \includegraphics[height=0.9\textheight]{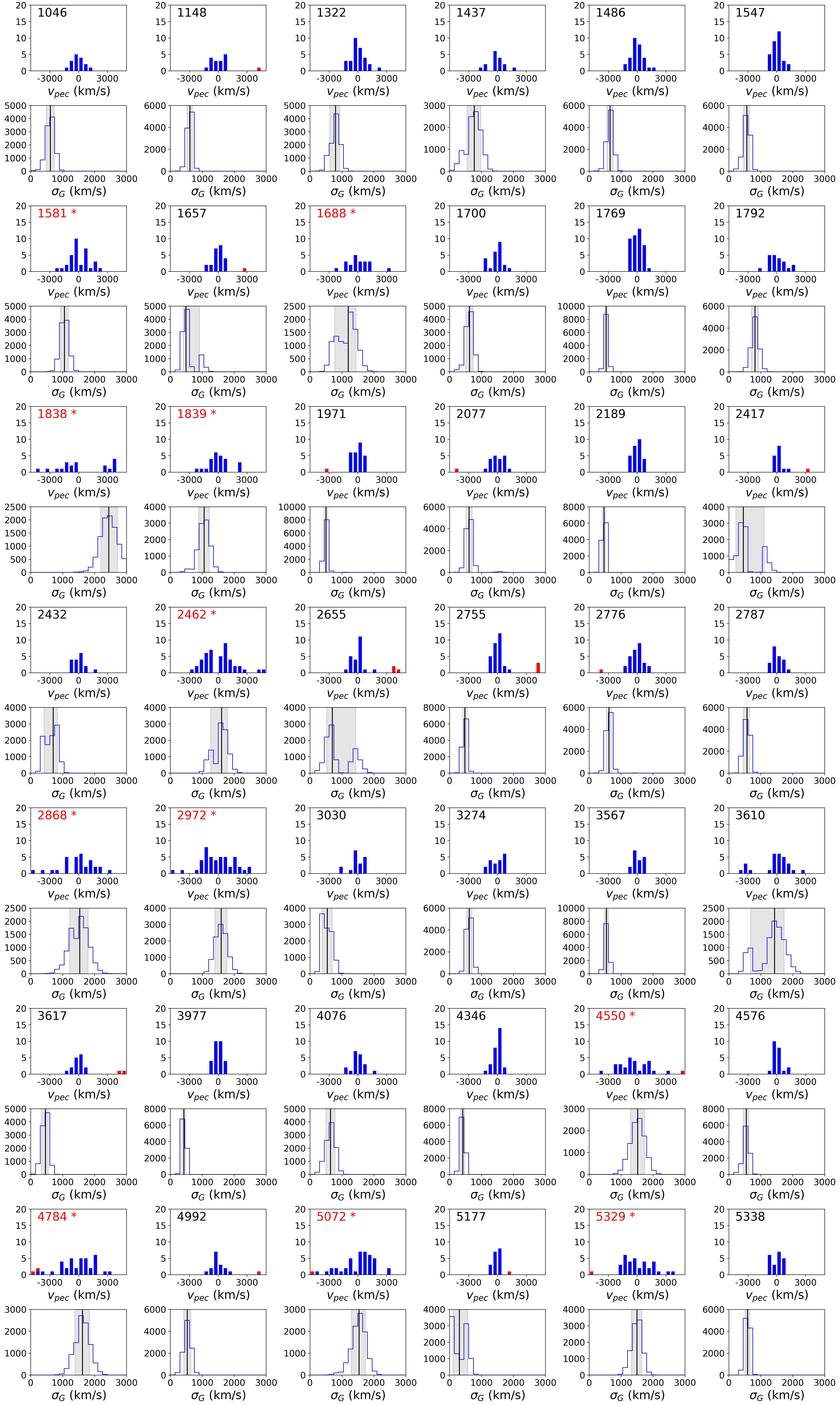}
        \caption{Gallery of all studied clusters. Cluster redMaPPer ID is shown in top left of each plot. Member galaxies are shown in blue, interlopers are shown in red. Outlier clusters are denoted with a red colored ID as well as an asterisk.}
    \end{subfigure}
\end{figure*}
\begin{figure*}
    \ContinuedFloat
    \begin{subfigure}{\textwidth}
        \centering
        \includegraphics[height=0.9\textheight]{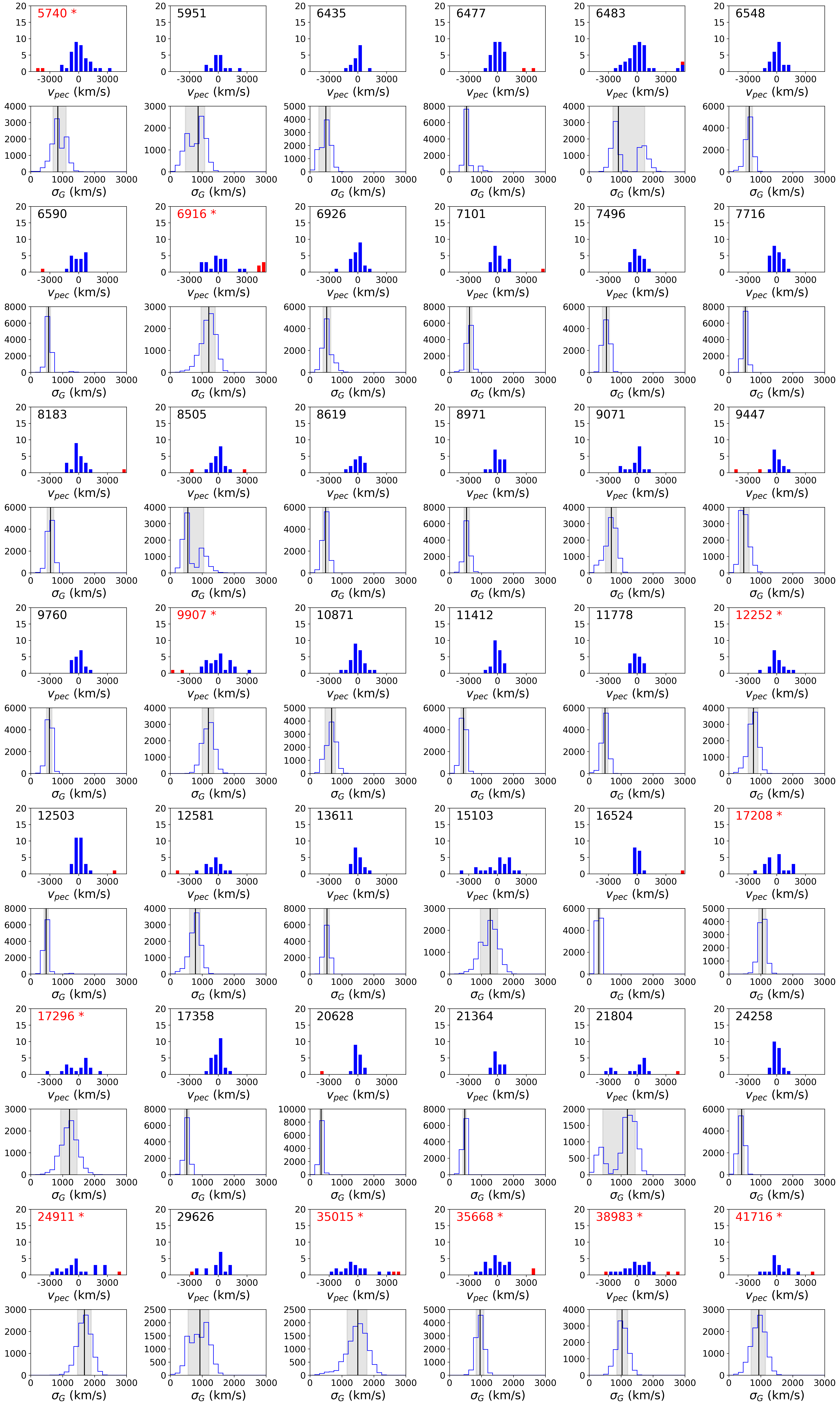}
        \caption{Gallery of all studied clusters. Cluster redMaPPer ID is shown in top left of each plot. Member galaxies are shown in blue, interlopers are shown in red. Outlier clusters are denoted with a red colored ID as well as an asterisk.}
    \end{subfigure}
\end{figure*}

\begin{figure}
    \centering
    \includegraphics[width=0.5\textwidth]{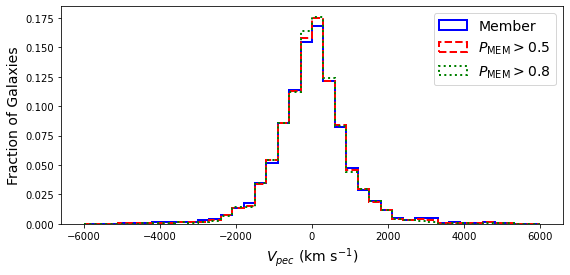}
    \caption{Histogram of the fractions of galaxies from all clusters stacked together in peculiar velocity bins for all member galaxies (solid blue), galaxies with $P_{\rm MEM}>0.5$ (dashed red), and galaxies with $P_{\rm MEM}>0.8$ (dotted green). Due to the similarity in the shape of the three histograms with differing $P_{\rm MEM}$ limits we decided not to add a $P_{\rm MEM}$ limit to our member selection process.}
    \label{fig:pmem_stacked}
\end{figure}

\begin{figure}
    \centering
    \includegraphics[width=0.5\textwidth]{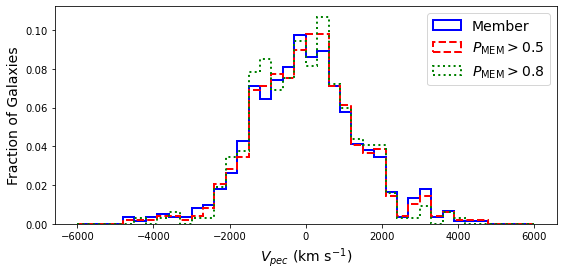}
    \caption{Histogram of the fractions of galaxies from outlier clusters stacked together in peculiar velocity bins for all member galaxies (solid blue, $\mu=-1$\kms, $\sigma=1437$\kms), galaxies with $P_{\rm MEM}>0.5$ (dashed red, $\mu=25$\kms, $\sigma=1391$\kms), and galaxies with $P_{\rm MEM}>0.8$ (dotted green, $\mu=-25$\kms, $\sigma=1304$\kms). These distributions are extremely similar, and limiting our sample based on $P_{\rm MEM}$ does not reduce or eliminate the outlier population.}
    \label{fig:pmem_outlier_stacked}
\end{figure}

\begin{figure}
    \centering
    \includegraphics[width=0.5\textwidth]{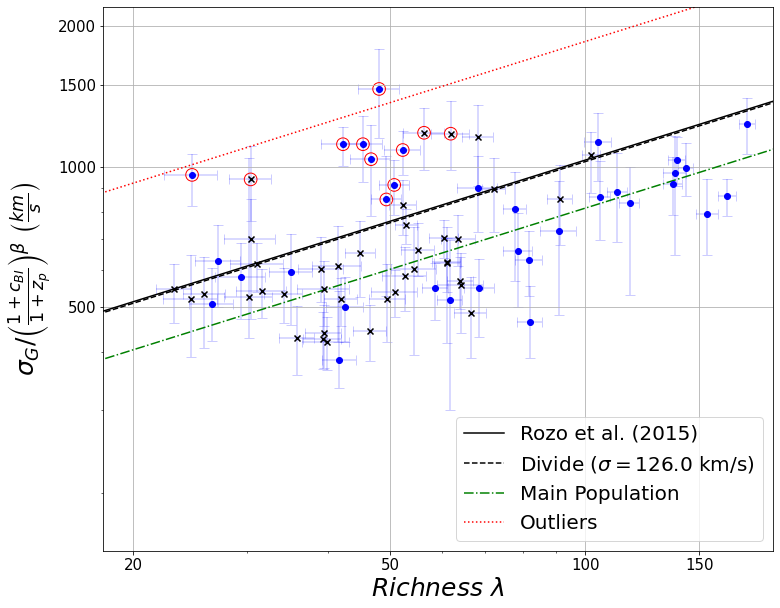}
    \caption{Figure \ref{fig:lamdba_sigma} for members selected using $P_{\rm MEM}>0.8$. While this appears to limit the outlier population it, is primarily due to many of the outlier clusters having fewer than 15 members with $P_{\rm MEM}>0.8$.}
    \label{fig:sig_g8_lam}
\end{figure}

Membership probability as determined by the redMaPPer algorithm was considered for interloper rejection but after stacking the galaxies from all of the clusters for different $P_{\rm MEM}$ limits (shown in Figure \ref{fig:pmem_stacked}) we observed little difference in the shape of the stacked histograms. Furthermore, we stacked all of the galaxies from the outlier population in Figure \ref{fig:pmem_outlier_stacked} for the same $P_{\rm MEM}$ limits which once again made little difference in the shape of the histograms. The similarity in the shapes of the stacked histograms suggests that applying any $P_{\rm MEM}$ limit to our interloper rejection would not measurably alter our velocity dispersions.  Indeed, we find little change in the overall $\sigma_G-\lambda$ relation when applying a $P_{\rm MEM}$ cut other than a reduction in the number of clusters that meet our criterion of having 15 spectroscopic members for fitting the velocity dispersion, which is shown in Figure \ref{fig:sig_g8_lam}. To be specific, cluster 648 enters the outlier population and cluster 1839 exits the outlier population. 

\begin{figure}
    \centering
    \includegraphics[width=0.5\textwidth]{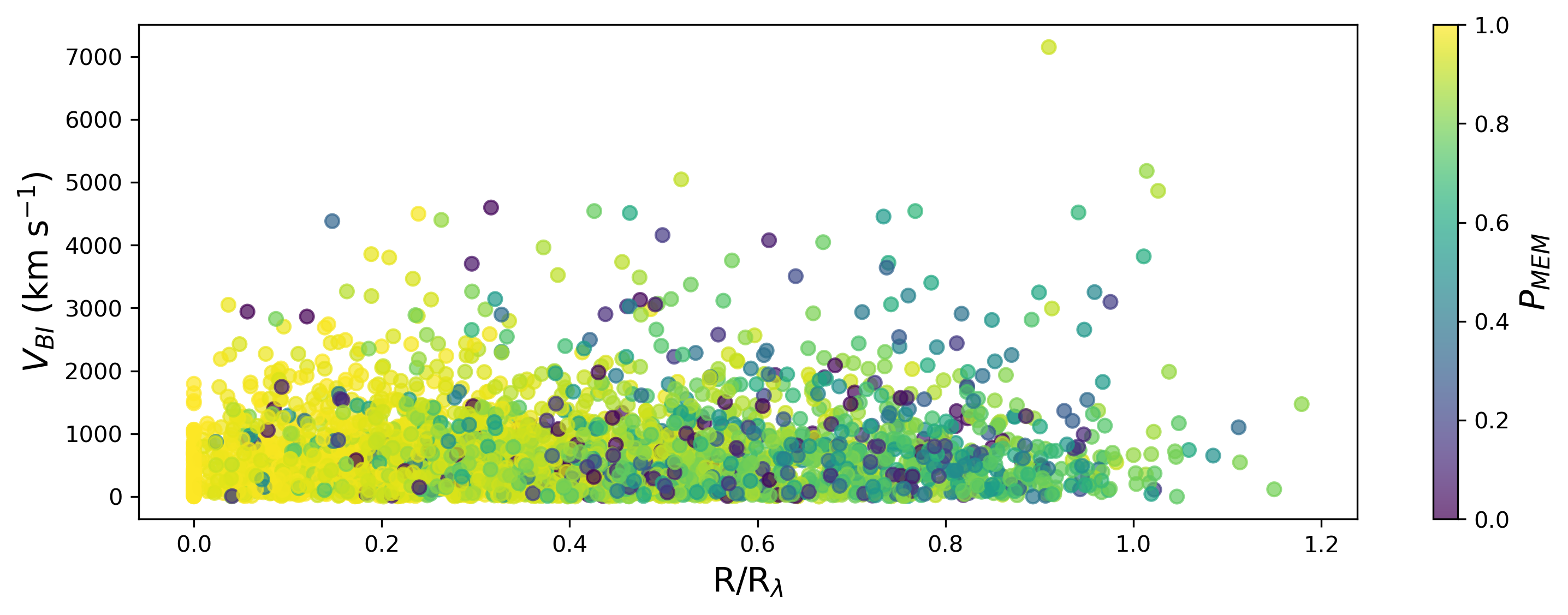}
    \caption{Member galaxy peculiar velocity shown against R/R$_\lambda$ with $P_{\rm MEM}$ on the color axis. The dependence of $P_{\rm MEM}$ on R/R$_\lambda$ is apparent with high $P_{\rm MEM}$ galaxies on average having low $R/R_\lambda$. }
    \label{fig:r_inv}
\end{figure}

Another potential method of interloper rejection is a cut on distance from the redMaPPer assigned center, $R/R(\lambda)$. However, we found this to be an ineffective way of limiting the outlier population. Figure \ref{fig:r_inv} shows a strong correlation between $R/R(\lambda)$ and $P_{\rm MEM}$ which is to be expected as $P_{\rm MEM}$ is dependant upon $R$. For this reason, a cut on $R/R(\lambda)$ yields a similar result to a cut on $P_{\rm MEM}$. This method of interloper rejection also does not account for miscentered clusters for which the $R$ values assigned to member galaxies are not representative of the galaxies position in relation to the cluster.

\begin{figure}
    \centering
    \includegraphics[width=0.5\textwidth]{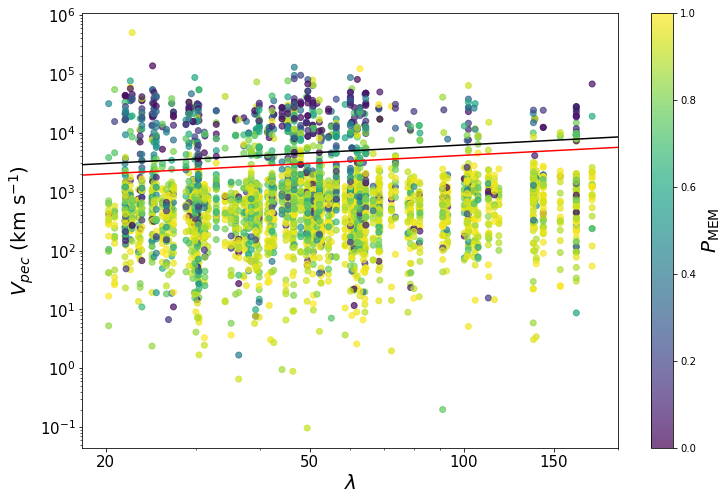}
    \includegraphics[width=0.5\textwidth]{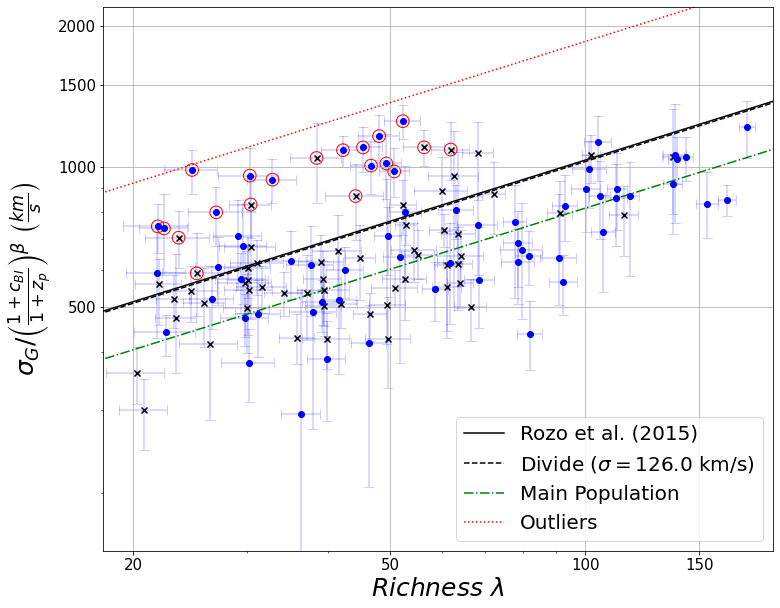}
    \caption{\textit{Top}: Equivalent to Figure \ref{fig:vel_offset} but showing a modified velocity offset limit in red. This modified limit changes the 3000 \kms in Equation \ref{eq:cut} to 2000 \kms.  \textit{Bottom}: Equivalent of Figure \ref{fig:lamdba_sigma} for members selected using the modified velocity offset limit.}
    \label{fig:2000kms_cut}
\end{figure}

In the process of better understanding the outlier population we also tested an altered cut on the initial galaxy sample considered as potential cluster members to see the effect on the cluster velocity dispersions.  Here we used a cut of
\begin{equation}
    |v|\leq(2000\textrm{ km s}^{-1})\left(\frac{\lambda}{20}\right)^{0.45}
    \label{eq:cut2000}
\end{equation}
lowering the normalization compared to Equation \ref{eq:cut}.
The difference between Equation \ref{eq:cut} and Equation \ref{eq:cut2000} is shown in Figure \ref{fig:2000kms_cut}. This resulted in a lower normalization for the outlier population which can be observed in Figure \ref{fig:2000kms_cut}. While this lower normalization does bring the outlier population closer to the main population in velocity dispersion it is still apparent in both the full sample and the redshift limited sample. This shows that the large velocity dispersions of the outlier clusters are not simply due to a small number of unrejected interloping galaxies.  A stricter interloper cut suppresses the velocity dispersions somewhat by artificially cutting off the velocity range but does not change the broad velocity distributions in these clusters.

\newpage

\section{DES Y3 Velocity Dispersion Sample}
Table \ref{tab:results} gives the catalog of redshift and velocity dispersion measurements for the clusters in our sample.  Listed are the redMaPPer MEM\_MATCH\_ID, number of members used for estimating the velocity dispersion, number of putative redMaPPer members cut, redMaPPer redshift, redMaPPer central galaxy redshift if available, the biweight location, the redMaPPer richness, the velocity dispersion estimated with the gapper method, and the velocity dispersion estimated with the biweight scale. Outlier clusters are denoted with an asterisk.

\renewcommand{\arraystretch}{1.3}

\begin{table*}
\begin{tabular}{|c|c|c|c|c|c|c|c|c|c|c|}
\hline
MEM$\_$MATCH$\_$ID & RA (J2000) & DEC (J2000) & $N_{\rm members}$ & $N_{\rm cut}$ & $z_\lambda$ & $z_{\rm center}$ & $z_{BI}$ & $\lambda$ & $\sigma_G$ (\kms) & $\sigma_{BI}$ (\kms) \\
\hline
6 & 04 11 11 & -48 19 40 & 22 & 2 & 0.413 & -- & 0.423 & 178$\pm$5 & $1351_{-200}^{+185}$ & $1375_{-158}^{+158}$ \\
11 & 04 16 09 & -24 04 03 & 81 & 23 & 0.391 & -- & 0.399 & 166$\pm$5 & $933_{-71}^{+69}$ & $967_{-75}^{+70}$ \\
19 & 00 40 50 & -44 07 53 & 42 & 3 & 0.361 & -- & 0.350 & 143$\pm$4 & $1136_{-127}^{+123}$ & $1147_{-115}^{+124}$ \\
32 & 02 31 41 & -04 52 57 & 60 & 9 & 0.192 & 0.185 & 0.187 & 102$\pm$3 & $1065_{-147}^{+107}$ & $1069_{-157}^{+116}$ \\
37 & 02 43 39 & -48 33 39 & 22 & 0 & 0.496 & -- & 0.499 & 139$\pm$4 & $1187_{-140}^{+134}$ & $1254_{-140}^{+138}$ \\
51 & 03 04 17 & -44 01 32 & 20 & 3 & 0.454 & -- & 0.458 & 138$\pm$6 & $1185_{-325}^{+351}$ & $1211_{-283}^{+445}$ \\
52 & 23 06 54 & -65 05 17 & 31 & 2 & 0.521 & 0.528 & 0.530 & 137$\pm$5 & $1061_{-141}^{+144}$ & $1091_{-153}^{+160}$ \\
54 & 23 35 08 & -45 44 21 & 21 & 3 & 0.550 & 0.547 & 0.547 & 154$\pm$6 & $971_{-157}^{+163}$ & $1041_{-283}^{+220}$ \\
64 & 21 59 59 & -62 45 14 & 16 & 0 & 0.386 & -- & 0.392 & 117$\pm$4 & $947_{-306}^{+176}$ & $821_{-214}^{+240}$ \\
71 & 02 30 55 & +02 47 20 & 19 & 1 & 0.239 & -- & 0.244 & 112$\pm$4 & $928_{-162}^{+179}$ & $993_{-197}^{+194}$ \\
81 & 01 23 11 & -48 21 23 & 16 & 3 & 0.639 & -- & 0.656 & 137$\pm$5 & $1274_{-407}^{+329}$ & $1224_{-355}^{+350}$ \\
86 & 01 52 42 & +01 00 25 & 22 & 3 & 0.232 & 0.230 & 0.231 & 105$\pm$4 & $883_{-168}^{+169}$ & $932_{-298}^{+215}$ \\
90 & 02 56 31 & +00 06 03 & 18 & 1 & 0.370 & 0.371 & 0.363 & 105$\pm$5 & $1220_{-210}^{+170}$ & $1185_{-234}^{+199}$ \\
151 & 04 17 23 & -47 48 48 & 17 & 0 & 0.590 & -- & 0.581 & 112$\pm$4 & $1687_{-850}^{+391}$ & $1162_{-213}^{+587}$ \\
184 & 02 48 08 & -02 16 37 & 18 & 1 & 0.237 & 0.234 & 0.237 & 93$\pm$4 & $850_{-139}^{+133}$ & $893_{-148}^{+157}$ \\
205 & 01 27 17 & +00 20 41 & 18 & 4 & 0.375 & 0.380 & 0.378 & 107$\pm$4 & $851_{-235}^{+792}$ & $792_{-238}^{+263}$ \\
225 & 00 34 28 & +02 25 23 & 17 & 1 & 0.392 & -- & 0.386 & 100$\pm$5 & $982_{-197}^{+186}$ & $1077_{-195}^{+215}$ \\
303 & 22 22 51 & -48 34 35 & 20 & 2 & 0.666 & -- & 0.653 & 91$\pm$4 & $956_{-180}^{+172}$ & $1039_{-152}^{+133}$ \\
321 & 01 08 03 & +02 51 60 & 17 & 1 & 0.326 & -- & 0.322 & 78$\pm$3 & $817_{-260}^{+162}$ & $722_{-232}^{+219}$ \\
340 & 22 33 16 & -53 39 09 & 16 & 0 & 0.430 & -- & 0.439 & 93$\pm$5 & $631_{-90}^{+91}$ & $674_{-91}^{+87}$ \\
381 & 00 44 28 & +01 50 11 & 17 & 2 & 0.371 & -- & 0.357 & 82$\pm$4 & $699_{-91}^{+90}$ & $765_{-91}^{+90}$ \\
398 & 04 06 55 & -48 04 57 & 15 & 0 & 0.732 & -- & 0.738 & 115$\pm$6 & $976_{-179}^{+152}$ & $1071_{-144}^{+123}$ \\
408 & 01 01 39 & +02 36 55 & 19 & 2 & 0.320 & 0.328 & 0.327 & 79$\pm$4 & $738_{-146}^{+124}$ & $728_{-123}^{+139}$ \\
414 & 00 08 10 & +02 01 13 & 20 & 3 & 0.367 & 0.365 & 0.366 & 82$\pm$4 & $474_{-78}^{+84}$ & $532_{-82}^{+96}$ \\
425 & 02 01 47 & -02 11 54 & 23 & 2 & 0.187 & 0.193 & 0.196 & 72$\pm$3 & $887_{-160}^{+141}$ & $925_{-179}^{+168}$ \\
500 & 00 17 38 & +00 52 42 & 21 & 5 & 0.210 & 0.212 & 0.213 & 62$\pm$3 & $656_{-173}^{+283}$ & $745_{-241}^{+1142}$ \\
513 & 00 23 01 & +00 09 17 & 38 & 4 & 0.154 & 0.158 & 0.158 & 64$\pm$3 & $561_{-68}^{+62}$ & $575_{-66}^{+62}$ \\
516 & 01 53 34 & -01 18 09 & 29 & 1 & 0.244 & 0.244 & 0.243 & 91$\pm$6 & $650_{-128}^{+252}$ & $640_{-136}^{+213}$ \\
550 & 01 22 03 & +00 20 04 & 30 & 4 & 0.176 & 0.175 & 0.175 & 61$\pm$3 & $616_{-184}^{+116}$ & $585_{-142}^{+119}$ \\
551 & 02 14 40 & -04 33 35 & 63 & 7 & 0.141 & -- & 0.140 & 60$\pm$3 & $722_{-59}^{+60}$ & $737_{-69}^{+63}$ \\
566 & 01 56 38 & +00 50 47 & 25 & 2 & 0.221 & -- & 0.218 & 59$\pm$3 & $559_{-81}^{+76}$ & $561_{-89}^{+82}$ \\
584 & 02 06 23 & -01 18 31 & 21 & 3 & 0.193 & 0.198 & 0.196 & 53$\pm$2 & $759_{-69}^{+65}$ & $795_{-71}^{+66}$ \\
607 & 02 12 27 & -05 37 35 & 19 & 4 & 0.309 & 0.300 & 0.299 & 69$\pm$4 & $605_{-130}^{+180}$ & $601_{-110}^{+259}$ \\
613 & 03 34 07 & -46 59 02 & 18 & 0 & 0.480 & -- & 0.486 & 53$\pm$3 & $909_{-267}^{+229}$ & $1065_{-294}^{+221}$ \\
640 & 00 34 23 & +00 51 26 & 30 & 1 & 0.188 & 0.192 & 0.190 & 64$\pm$4 & $720_{-95}^{+364}$ & $750_{-91}^{+119}$ \\
648 & 02 02 02 & +03 44 51 & 19 & 0 & 0.164 & -- & 0.164 & 68$\pm$4 & $1063_{-212}^{+176}$ & $1031_{-187}^{+196}$ \\
658 & 05 42 50 & -41 00 00 & 19 & 2 & 0.654 & -- & 0.640 & 101$\pm$6 & $1184_{-202}^{+182}$ & $1167_{-235}^{+204}$ \\
745 & 02 45 52 & +00 42 16 & 36 & 2 & 0.178 & 0.180 & 0.181 & 61$\pm$3 & $553_{-106}^{+75}$ & $537_{-104}^{+83}$ \\
761 & 21 46 06 & -48 46 53 & 19 & 1 & 0.625 & -- & 0.623 & 79$\pm$5 & $744_{-138}^{+131}$ & $777_{-164}^{+145}$ \\
812 & 02 10 08 & +02 54 27 & 24 & 0 & 0.148 & 0.152 & 0.148 & 52$\pm$3 & $819_{-110}^{+106}$ & $846_{-103}^{+99}$ \\
844 & 01 31 26 & -04 44 59 & 16 & 1 & 0.217 & 0.217 & 0.217 & 50$\pm$2 & $726_{-124}^{+113}$ & $749_{-131}^{+124}$ \\
992 & 21 35 40 & +00 09 57 & 23 & 8 & 0.118 & -- & 0.119 & 55$\pm$3 & $648_{-82}^{+175}$ & $689_{-84}^{+91}$ \\
1046 & 01 58 26 & -01 46 39 & 16 & 1 & 0.157 & 0.163 & 0.163 & 64$\pm$3 & $616_{-123}^{+125}$ & $679_{-154}^{+125}$ \\
1148 & 01 06 33 & -02 27 02 & 16 & 1 & 0.191 & 0.186 & 0.189 & 30$\pm$3 & $613_{-85}^{+74}$ & $650_{-83}^{+76}$ \\
1322 & 03 40 07 & -28 50 38 & 30 & 4 & 0.338 & 0.336 & 0.337 & 68$\pm$5 & $804_{-186}^{+125}$ & $776_{-170}^{+138}$ \\
1437 & 04 56 28 & -51 16 35 & 16 & 1 & 0.565 & 0.562 & 0.562 & 80$\pm$6 & $773_{-228}^{+203}$ & $778_{-421}^{+231}$ \\
\hline

\multicolumn{8}{l}{$^*$ indicates outlier clusters}
\end{tabular}
\end{table*}

\begin{table*}
\begin{tabular}{|c|c|c|c|c|c|c|c|c|c|c|}
\hline
MEM$\_$MATCH$\_$ID & RA (J2000) & DEC (J2000) & $N_{\rm members}$ & $N_{\rm cut}$ & $z_\lambda$ & $z_{center}$ & $z_{BI}$ & $\lambda$ & $\sigma_G$ (\kms) & $\sigma_{BI}$ (\kms) \\
\hline
1486 & 21 25 46 & +00 55 52 & 30 & 4 & 0.127 & 0.135 & 0.136 & 54$\pm$4 & $653_{-96}^{+94}$ & $659_{-110}^{+105}$ \\
1547 & 02 25 45 & -03 12 33 & 31 & 3 & 0.141 & 0.142 & 0.141 & 53$\pm$4 & $565_{-101}^{+85}$ & $535_{-132}^{+110}$ \\
1581* & 02 15 28 & -04 40 41 & 33 & 10 & 0.352 & 0.348 & 0.352 & 51$\pm$3 & $1057_{-124}^{+125}$ & $1084_{-141}^{+140}$ \\
1657 & 01 39 16 & -03 38 04 & 23 & 5 & 0.115 & -- & 0.115 & 49$\pm$3 & $492_{-86}^{+424}$ & $527_{-87}^{+84}$ \\
1688* & 01 31 20 & -13 28 15 & 21 & 3 & 0.214 & -- & 0.210 & 27$\pm$3 & $1199_{-424}^{+245}$ & $1032_{-209}^{+329}$ \\
1700 & 02 47 03 & +04 23 21 & 23 & 1 & 0.137 & -- & 0.140 & 45$\pm$2 & $625_{-114}^{+112}$ & $656_{-129}^{+105}$ \\
1769 & 03 36 51 & -28 04 44 & 43 & 11 & 0.120 & 0.105 & 0.105 & 51$\pm$4 & $533_{-48}^{+49}$ & $541_{-46}^{+45}$ \\
1792 & 00 20 16 & +00 04 46 & 21 & 0 & 0.201 & 0.212 & 0.211 & 63$\pm$4 & $822_{-114}^{+111}$ & $851_{-140}^{+141}$ \\
1838* & 22 14 52 & +01 44 39 & 19 & 3 & 0.691 & 0.683 & 0.689 & 74$\pm$4 & $2446_{-252}^{+278}$ & $2699_{-280}^{+251}$ \\
1839* & 01 06 50 & +01 03 56 & 25 & 5 & 0.253 & -- & 0.254 & 49$\pm$3 & $1058_{-174}^{+165}$ & $1067_{-221}^{+221}$ \\
1971 & 01 02 45 & +01 07 60 & 26 & 1 & 0.149 & -- & 0.144 & 42$\pm$2 & $502_{-55}^{+52}$ & $518_{-52}^{+51}$ \\
2077 & 03 10 32 & -46 47 02 & 20 & 1 & 0.708 & -- & 0.706 & 67$\pm$4 & $614_{-96}^{+93}$ & $656_{-88}^{+92}$ \\
2189 & 01 48 28 & -04 07 47 & 27 & 1 & 0.108 & 0.086 & 0.087 & 47$\pm$3 & $464_{-55}^{+50}$ & $481_{-58}^{+51}$ \\
2417 & 01 52 06 & +01 32 39 & 15 & 3 & 0.217 & -- & 0.215 & 46$\pm$3 & $456_{-236}^{+652}$ & $346_{-152}^{+202}$ \\
2432 & 00 32 18 & +01 00 38 & 17 & 1 & 0.381 & 0.390 & 0.387 & 52$\pm$3 & $705_{-287}^{+134}$ & $493_{-71}^{+272}$ \\
2462* & 03 34 15 & -28 26 49 & 45 & 15 & 0.651 & 0.657 & 0.660 & 56$\pm$4 & $1608_{-344}^{+179}$ & $1557_{-291}^{+204}$ \\
2655 & 00 45 50 & +00 51 01 & 23 & 9 & 0.110 & 0.111 & 0.110 & 41$\pm$4 & $700_{-173}^{+739}$ & $652_{-130}^{+620}$ \\
2755 & 01 31 33 & +00 33 22 & 29 & 13 & 0.103 & 0.079 & 0.080 & 39$\pm$2 & $480_{-65}^{+62}$ & $480_{-57}^{+61}$ \\
2776 & 02 43 12 & -01 01 12 & 28 & 6 & 0.240 & 0.239 & 0.240 & 43$\pm$3 & $619_{-80}^{+74}$ & $636_{-84}^{+76}$ \\
2787 & 21 30 27 & +00 00 24 & 21 & 1 & 0.133 & 0.137 & 0.135 & 39$\pm$3 & $568_{-118}^{+80}$ & $543_{-81}^{+89}$ \\
2868* & 03 33 59 & -28 38 11 & 31 & 13 & 0.657 & 0.664 & 0.663 & 62$\pm$4 & $1537_{-319}^{+261}$ & $1632_{-373}^{+299}$ \\
2972* & 02 16 36 & -04 27 05 & 46 & 7 & 0.443 & 0.448 & 0.448 & 52$\pm$3 & $1593_{-198}^{+182}$ & $1649_{-220}^{+196}$ \\
3030 & 02 15 30 & -05 32 55 & 18 & 8 & 0.287 & 0.290 & 0.290 & 42$\pm$3 & $542_{-162}^{+150}$ & $624_{-213}^{+185}$ \\
3274 & 01 56 54 & -04 24 26 & 19 & 5 & 0.136 & 0.134 & 0.135 & 39$\pm$3 & $614_{-80}^{+87}$ & $674_{-86}^{+86}$ \\
3567 & 00 44 37 & +00 55 20 & 18 & 2 & 0.202 & 0.201 & 0.197 & 34$\pm$2 & $542_{-57}^{+62}$ & $577_{-70}^{+61}$ \\
3610 & 03 29 31 & -28 20 09 & 28 & 8 & 0.678 & 0.001 & 0.680 & 63$\pm$6 & $1435_{-755}^{+298}$ & $1551_{-890}^{+322}$ \\
3617 & 02 28 29 & -04 43 43 & 16 & 6 & 0.611 & 0.612 & 0.611 & 40$\pm$3 & $460_{-124}^{+96}$ & $535_{-99}^{+93}$ \\
3977 & 03 32 27 & -27 29 39 & 28 & 6 & 0.158 & 0.148 & 0.147 & 36$\pm$2 & $424_{-52}^{+55}$ & $450_{-55}^{+56}$ \\
4076 & 02 49 12 & +00 48 49 & 20 & 1 & 0.269 & 0.272 & 0.271 & 38$\pm$3 & $645_{-141}^{+139}$ & $658_{-139}^{+141}$ \\
4346 & 00 21 42 & +00 52 32 & 28 & 14 & 0.108 & 0.105 & 0.106 & 40$\pm$3 & $413_{-59}^{+56}$ & $454_{-56}^{+55}$ \\
4550* & 02 23 58 & -04 35 05 & 28 & 21 & 0.492 & 0.494 & 0.497 & 45$\pm$3 & $1518_{-226}^{+216}$ & $1517_{-210}^{+223}$ \\
4576 & 02 13 56 & -01 31 19 & 22 & 0 & 0.169 & 0.173 & 0.176 & 39$\pm$4 & $547_{-92}^{+80}$ & $516_{-151}^{+119}$ \\
4784* & 00 34 42 & -43 50 39 & 36 & 17 & 0.542 & 0.553 & 0.545 & 48$\pm$3 & $1622_{-237}^{+225}$ & $1700_{-222}^{+241}$ \\
4992 & 02 45 01 & -03 05 54 & 16 & 5 & 0.161 & 0.162 & 0.162 & 37$\pm$3 & $531_{-111}^{+111}$ & $555_{-124}^{+118}$ \\
5072* & 02 17 35 & -05 13 30 & 42 & 33 & 0.643 & 0.648 & 0.643 & 47$\pm$3 & $1538_{-242}^{+202}$ & $1625_{-223}^{+223}$ \\
5177 & 02 23 33 & -07 13 40 & 18 & 2 & 0.274 & 0.279 & 0.280 & 36$\pm$3 & $307_{-202}^{+264}$ & $315_{-206}^{+73}$ \\
5329* & 02 23 51 & -05 36 40 & 33 & 8 & 0.490 & 0.498 & 0.500 & 42$\pm$3 & $1495_{-175}^{+150}$ & $1450_{-161}^{+168}$ \\
5338 & 02 03 02 & -04 59 38 & 21 & 0 & 0.494 & 0.512 & 0.509 & 39$\pm$3 & $587_{-79}^{+71}$ & $608_{-73}^{+64}$ \\
5740* & 02 16 12 & -04 14 22 & 36 & 6 & 0.154 & 0.153 & 0.153 & 30$\pm$2 & $850_{-151}^{+260}$ & $815_{-284}^{+259}$ \\
5951 & 00 47 31 & +00 52 57 & 16 & 3 & 0.117 & 0.117 & 0.119 & 30$\pm$2 & $871_{-396}^{+208}$ & $643_{-156}^{+339}$ \\
6435 & 01 44 54 & -02 17 05 & 16 & 5 & 0.235 & 0.237 & 0.237 & 38$\pm$3 & $500_{-220}^{+136}$ & $524_{-216}^{+156}$ \\
6477 & 00 46 24 & +00 00 09 & 30 & 3 & 0.117 & 0.116 & 0.114 & 30$\pm$3 & $533_{-61}^{+67}$ & $564_{-56}^{+57}$ \\
6483 & 00 36 45 & -44 10 50 & 40 & 19 & 0.870 & 0.871 & 0.870 & 64$\pm$5 & $915_{-171}^{+824}$ & $887_{-125}^{+812}$ \\
6548 & 02 01 17 & -01 24 31 & 23 & 0 & 0.212 & 0.209 & 0.209 & 35$\pm$3 & $640_{-109}^{+102}$ & $639_{-140}^{+132}$ \\
6590 & 01 04 59 & -02 42 02 & 20 & 1 & 0.195 & 0.192 & 0.189 & 32$\pm$3 & $558_{-69}^{+64}$ & $598_{-65}^{+73}$ \\
\hline

\multicolumn{8}{l}{$^*$ indicates outlier clusters}
\end{tabular}
\end{table*}

\begin{table*}
\begin{tabular}{|c|c|c|c|c|c|c|c|c|c|c|}
\hline
MEM$\_$MATCH$\_$ID & RA (J2000) & DEC (J2000) & $N_{\rm members}$ & $N_{\rm cut}$ & $z_\lambda$ & $z_{center}$ & $z_{BI}$ & $\lambda$ & $\sigma_G$ (\kms) & $\sigma_{BI}$ (\kms) \\
\hline
6916* & 00 03 49 & +02 02 56 & 22 & 5 & 0.109 & -- & 0.096 & 38$\pm$3 & $1206_{-245}^{+201}$ & $1140_{-242}^{+248}$ \\
6926 & 02 28 30 & +00 30 36 & 23 & 13 & 0.733 & -- & 0.721 & 50$\pm$4 & $526_{-113}^{+128}$ & $531_{-91}^{+241}$ \\
7101 & 02 35 12 & -01 30 47 & 21 & 2 & 0.170 & 0.173 & 0.173 & 31$\pm$3 & $624_{-90}^{+74}$ & $638_{-145}^{+88}$ \\
7496 & 00 35 40 & +01 37 42 & 20 & 1 & 0.102 & -- & 0.080 & 30$\pm$3 & $540_{-125}^{+93}$ & $522_{-86}^{+98}$ \\
7716 & 01 59 31 & +00 06 16 & 24 & 1 & 0.155 & 0.156 & 0.156 & 23$\pm$1 & $517_{-69}^{+63}$ & $517_{-61}^{+59}$ \\
8183 & 02 25 12 & -06 22 59 & 22 & 7 & 0.209 & 0.204 & 0.204 & 27$\pm$2 & $620_{-103}^{+100}$ & $668_{-127}^{+102}$ \\
8505 & 01 32 47 & +01 15 46 & 20 & 5 & 0.123 & 0.126 & 0.125 & 26$\pm$2 & $550_{-132}^{+497}$ & $525_{-104}^{+173}$ \\
8619 & 02 02 10 & -03 11 15 & 15 & 2 & 0.153 & 0.154 & 0.153 & 30$\pm$2 & $491_{-96}^{+91}$ & $542_{-107}^{+89}$ \\
8971 & 02 01 46 & -01 40 13 & 17 & 0 & 0.205 & 0.209 & 0.208 & 27$\pm$2 & $531_{-75}^{+95}$ & $578_{-103}^{+119}$ \\
9071 & 02 25 49 & -05 53 46 & 17 & 6 & 0.243 & 0.233 & 0.232 & 30$\pm$2 & $694_{-185}^{+160}$ & $784_{-183}^{+137}$ \\
9447 & 02 11 03 & -04 53 38 & 15 & 7 & 0.137 & 0.138 & 0.138 & 23$\pm$2 & $468_{-114}^{+171}$ & $473_{-118}^{+245}$ \\
9760 & 00 32 11 & +00 39 60 & 19 & 3 & 0.206 & 0.215 & 0.215 & 29$\pm$3 & $585_{-89}^{+81}$ & $593_{-102}^{+89}$ \\
9907* & 02 15 36 & -04 00 41 & 27 & 9 & 0.373 & 0.383 & 0.376 & 33$\pm$3 & $1193_{-196}^{+167}$ & $1180_{-144}^{+159}$ \\
10871 & 01 12 04 & +00 43 52 & 27 & 1 & 0.174 & 0.179 & 0.179 & 30$\pm$3 & $678_{-205}^{+124}$ & $648_{-199}^{+152}$ \\
11412 & 02 24 29 & -04 49 14 & 23 & 6 & 0.485 & 0.495 & 0.495 & 30$\pm$3 & $435_{-77}^{+80}$ & $441_{-118}^{+98}$ \\
11778 & 02 10 18 & -03 09 55 & 17 & 3 & 0.246 & 0.245 & 0.244 & 31$\pm$3 & $495_{-94}^{+95}$ & $528_{-152}^{+122}$ \\
12252* & 02 19 56 & -05 28 03 & 18 & 7 & 0.278 & 0.279 & 0.278 & 22$\pm$2 & $771_{-160}^{+143}$ & $773_{-179}^{+170}$ \\
12503 & 03 29 27 & -27 31 26 & 29 & 3 & 0.236 & 0.219 & 0.218 & 30$\pm$3 & $485_{-65}^{+66}$ & $487_{-74}^{+71}$ \\
12581 & 00 38 48 & -43 49 13 & 16 & 15 & 0.413 & 0.403 & 0.401 & 29$\pm$2 & $785_{-181}^{+151}$ & $830_{-231}^{+171}$ \\
13611 & 01 34 54 & +00 39 53 & 19 & 2 & 0.103 & 0.084 & 0.082 & 22$\pm$2 & $535_{-99}^{+74}$ & $511_{-74}^{+81}$ \\
15103 & 02 23 43 & -05 02 01 & 23 & 19 & 0.869 & 0.859 & 0.854 & 60$\pm$6 & $1272_{-307}^{+235}$ & $1423_{-273}^{+241}$ \\
16524 & 02 33 53 & +00 04 40 & 16 & 3 & 0.184 & 0.186 & 0.186 & 21$\pm$2 & $302_{-54}^{+51}$ & $310_{-49}^{+44}$ \\
17208* & 02 22 05 & -04 33 00 & 20 & 8 & 0.317 & 0.319 & 0.317 & 25$\pm$2 & $1052_{-117}^{+106}$ & $1111_{-133}^{+143}$ \\
17296* & 02 30 25 & +00 37 43 & 18 & 8 & 0.824 & -- & 0.863 & 44$\pm$5 & $1218_{-278}^{+235}$ & $1337_{-230}^{+253}$ \\
17358 & 23 35 28 & +01 02 48 & 26 & 4 & 0.106 & 0.084 & 0.084 & 25$\pm$2 & $518_{-74}^{+71}$ & $536_{-74}^{+67}$ \\
20628 & 22 56 28 & +00 32 54 & 18 & 2 & 0.111 & 0.110 & 0.110 & 20$\pm$2 & $351_{-49}^{+48}$ & $366_{-61}^{+56}$ \\
21364 & 03 30 06 & -28 01 56 & 15 & 6 & 0.344 & 0.337 & 0.339 & 23$\pm$2 & $474_{-56}^{+43}$ & $483_{-72}^{+52}$ \\
21804 & 22 04 43 & +01 13 12 & 15 & 5 & 0.564 & 0.554 & 0.552 & 29$\pm$3 & $1196_{-770}^{+248}$ & $1483_{-369}^{+182}$ \\
24258 & 22 35 12 & -01 08 50 & 23 & 2 & 0.109 & 0.090 & 0.090 & 26$\pm$3 & $400_{-123}^{+79}$ & $387_{-128}^{+96}$ \\
24911* & 03 27 59 & -29 06 35 & 22 & 5 & 0.622 & -- & 0.606 & 30$\pm$3 & $1683_{-215}^{+210}$ & $1771_{-264}^{+209}$ \\
29626 & 02 23 11 & -04 12 52 & 18 & 5 & 0.628 & 0.625 & 0.630 & 22$\pm$2 & $928_{-369}^{+281}$ & $1147_{-387}^{+248}$ \\
35015* & 02 18 08 & -05 46 02 & 19 & 12 & 0.690 & 0.692 & 0.689 & 25$\pm$3 & $1501_{-337}^{+284}$ & $1369_{-366}^{+392}$ \\
35668* & 02 18 24 & -05 25 01 & 25 & 10 & 0.652 & 0.648 & 0.642 & 22$\pm$3 & $954_{-121}^{+119}$ & $1014_{-123}^{+116}$ \\
38983* & 03 29 04 & -29 05 50 & 22 & 15 & 0.723 & 0.720 & 0.711 & 24$\pm$3 & $1030_{-164}^{+174}$ & $1126_{-169}^{+163}$ \\
41716* & 02 17 54 & -05 27 06 & 16 & 13 & 0.679 & 0.691 & 0.692 & 25$\pm$3 & $939_{-241}^{+209}$ & $934_{-271}^{+224}$ \\
\hline
\multicolumn{8}{l}{$^*$ indicates outlier clusters}
\end{tabular}
\caption{Catalog of cluster measurements. Column 1 lists the redMaPPer MEM\_MATCH\_ID, columns 2 and 3 list the RA and DEC respectively, column 4 the number of members used for estimating the velocity dispersion, column 5 number of putative redMaPPer members cut, column 6 the redMaPPer redshift, column 7 the redMaPPer central galaxy redshift if available, column 8 the biweight location, column 9 the redMaPPer richness, column 10 the velocity dispersion estimated with the gapper method, and column 11 the velocity dispersion estimated with the biweight scale.}
\label{tab:results}
\end{table*}

\section{Affiliations}
$^{1}$Department of Physics, University of California, 1156 High St, Santa Cruz, CA 95064, USA\\
$^{2}$Santa Cruz Institute for Particle Physics, University of California, 1156 High St, Santa Cruz, CA 95064, USA\\
$^{3}$Department of Physics and Astronomy, University of Pennsylvania, 209 South 33rd Street, Philadelphia, PA 19104, USA\\
$^{4}$Department of Physics and Astronomy, Pevensey Building, University of Sussex, Brighton, BN1 9QH, UK\\
$^{5}$The Michigan Institute for Data Science, University of Michigan, Ann Arbor, MI 48109, USA\\
$^{6}$Department of Physics and Leinweber Center for Theoretical Physics, University of Michigan, Ann Arbor, MI 48109, USA\\
$^{7}$ Astronomy Unit, Department of Physics, University of Trieste, via Tiepolo 11, I-34131 Trieste, Italy\\
$^{8}$ IFPU - Institute for Fundamental Physics of the Universe, Via Beirut 2, 34014 Trieste, Italy\\
$^{9}$ INAF - Osservatorio Astronomico di Trieste, via G. B. Tiepolo 11, I-34143 Trieste, Italy\\
$^{10}$ INFN - National Institute for Nuclear Physics, Via Valerio 2, I-34127 Trieste, Italy\\
$^{11}$ Department of Physics, Stanford University, 382 Via Pueblo Mall, Stanford, CA 94305, USA\\
$^{12}$ Kavli Institute for Particle Astrophysics \& Cosmology, P. O. Box 2450, Stanford University, Stanford, CA 94305, USA\\
$^{13}$ SLAC National Accelerator Laboratory, Menlo Park, CA 94025, USA\\
$^{14}$ Astrophysics Research Institute, Liverpool John Moores University, Liverpool Science Park, 146 Brownlow Hill, Liverpool L3 5RF, UK\\
$^{15}$School of Mathematics, Statistics, and Computer Science, University of KwaZulu-Natal, Westville Campus, Durban 4041, SA\\
$^{16}$ Instituto de Astrofisica e Ciencias do Espaco, Faculdade de Ciencias, Universidade de Lisboa, 1769-016 Lisboa, Portugal\\
$^{17}$George P. and Cynthia Woods Mitchell Institute for Fundamental Physics and Astronomy, and Department of Physics and Astronomy, Texas A\&M University, College Station, TX 77843, USA\\
$^{18}$ Department of Physics and Astronomy, Clemson University, Kinard Lab of Physics, Clemson, SC 29634-0978, USA\\
$^{19}$Instituto de Física Gleb Wataghin, Universidade Estadual de Campinas, 13083-859, Campinas, SP, Brazil\\
$^{20}$Computer Science and Mathematics Division, Oak Ridge National Laboratory, Oak Ridge, TN 37831\\
$^{21}$Instituto de Astrofísica e Ciencias do Espaco, Universidade do Porto, CAUP, Rua das Estrelas, P-4150-762 Porto, Portugal\\
$^{22}$ Argonne National Laboratory, 9700 South Cass Avenue, Lemont, IL 60439, USA\\
$^{23}$ Cerro Tololo Inter-American Observatory, NSF's National Optical-Infrared Astronomy Research Laboratory, Casilla 603, La Serena, Chile\\
$^{24}$ Laborat\'orio Interinstitucional de e-Astronomia - LIneA, Rua Gal. Jos\'e Cristino 77, Rio de Janeiro, RJ - 20921-400, Brazil\\
$^{25}$ Fermi National Accelerator Laboratory, P. O. Box 500, Batavia, IL 60510, USA\\
$^{26}$ Instituto de F\'{i}sica Te\'orica, Universidade Estadual Paulista, S\~ao Paulo, Brazil\\
$^{27}$ Centro de Investigaciones Energ\'eticas, Medioambientales y Tecnol\'ogicas (CIEMAT), Madrid, Spain\\
$^{28}$ CNRS, UMR 7095, Institut d'Astrophysique de Paris, F-75014, Paris, France\\
$^{29}$ Sorbonne Universit\'es, UPMC Univ Paris 06, UMR 7095, Institut d'Astrophysique de Paris, F-75014, Paris, France\\
$^{30}$ School of Mathematics and Physics, University of Queensland,  Brisbane, QLD 4072, Australia\\
$^{31}$ Instituto de Astrofisica de Canarias, E-38205 La Laguna, Tenerife, Spain\\
$^{32}$ Universidad de La Laguna, Dpto. Astrofísica, E-38206 La Laguna, Tenerife, Spain\\
$^{33}$ INAF, Astrophysical Observatory of Turin, I-10025 Pino Torinese, Italy\\
$^{34}$ Center for Astrophysical Surveys, National Center for Supercomputing Applications, 1205 West Clark St., Urbana, IL 61801, USA\\
$^{35}$ Department of Astronomy, University of Illinois at Urbana-Champaign, 1002 W. Green Street, Urbana, IL 61801, USA\\
$^{36}$ Institut de F\'{\i}sica d'Altes Energies (IFAE), The Barcelona Institute of Science and Technology, Campus UAB, 08193 Bellaterra (Barcelona) Spain\\
$^{37}$ Center for Cosmology and Astro-Particle Physics, The Ohio State University, Columbus, OH 43210, USA\\
$^{38}$ Institut d'Estudis Espacials de Catalunya (IEEC), 08034 Barcelona, Spain\\
$^{39}$ Institute of Space Sciences (ICE, CSIC),  Campus UAB, Carrer de Can Magrans, s/n,  08193 Barcelona, Spain\\
$^{40}$ Observat\'orio Nacional, Rua Gal. Jos\'e Cristino 77, Rio de Janeiro, RJ - 20921-400, Brazil\\
$^{41}$ Department of Physics, University of Michigan, Ann Arbor, MI 48109, USA\\
$^{42}$ Department of Physics, IIT Hyderabad, Kandi, Telangana 502285, India\\
$^{43}$ Faculty of Physics, Ludwig-Maximilians-Universit\"at, Scheinerstr. 1, 81679 Munich, Germany\\
$^{44}$ Department of Physics \& Astronomy, University College London, Gower Street, London, WC1E 6BT, UK\\
$^{45}$ Department of Astronomy, University of Michigan, Ann Arbor, MI 48109, USA\\
$^{46}$ Institute of Theoretical Astrophysics, University of Oslo. P.O. Box 1029 Blindern, NO-0315 Oslo, Norway\\
$^{47}$ Kavli Institute for Cosmological Physics, University of Chicago, Chicago, IL 60637, USA\\
$^{48}$ Instituto de Fisica Teorica UAM/CSIC, Universidad Autonoma de Madrid, 28049 Madrid, Spain\\
$^{49}$ Centre for Astrophysics \& Supercomputing, Swinburne University of Technology, Victoria 3122, Australia\\
$^{50}$ Department of Physics, The Ohio State University, Columbus, OH 43210, USA\\
$^{51}$ Center for Astrophysics $\vert$ Harvard \& Smithsonian, 60 Garden Street, Cambridge, MA 02138, USA\\
$^{52}$ Australian Astronomical Optics, Macquarie University, North Ryde, NSW 2113, Australia\\
$^{53}$ Lowell Observatory, 1400 Mars Hill Rd, Flagstaff, AZ 86001, USA\\
$^{54}$ Sydney Institute for Astronomy, School of Physics, A28, The University of Sydney, NSW 2006, Australia\\
$^{55}$ Centre for Gravitational Astrophysics, College of Science, The Australian National University, ACT 2601, Australia\\
$^{56}$ The Research School of Astronomy and Astrophysics, Australian National University, ACT 2601, Australia\\
$^{57}$ Departamento de F\'isica Matem\'atica, Instituto de F\'isica, Universidade de S\~ao Paulo, CP 66318, S\~ao Paulo, SP, 05314-970, Brazil\\
$^{58}$ Instituci\'o Catalana de Recerca i Estudis Avan\c{c}ats, E-08010 Barcelona, Spain\\
$^{59}$ Physics Department, 2320 Chamberlin Hall, University of Wisconsin-Madison, 1150 University Avenue Madison, WI  53706-1390\\
$^{60}$ Institute of Astronomy, University of Cambridge, Madingley Road, Cambridge CB3 0HA, UK\\
$^{61}$ Department of Astrophysical Sciences, Princeton University, Peyton Hall, Princeton, NJ 08544, USA\\
$^{62}$ School of Physics and Astronomy, University of Southampton,  Southampton, SO17 1BJ, UK\\
$^{63}$ Institute of Cosmology and Gravitation, University of Portsmouth, Portsmouth, PO1 3FX, UK\\
$^{64}$ Max Planck Institute for Extraterrestrial Physics, Giessenbachstrasse, 85748 Garching, Germany\\
$^{65}$ Universit\"ats-Sternwarte, Fakult\"at f\"ur Physik, Ludwig-Maximilians Universit\"at M\"unchen, Scheinerstr. 1, 81679 M\"unchen, Germany\\
$^{66}$ Departamento de Física e Astronomia, Faculdade de Ciências, Universidade do Porto, Rua do Campo Alegre, 687, 4169-007 Porto, Portugal

\bsp	
\label{lastpage}
\end{document}